\documentclass[a4paper, 12pt]{article}

\usepackage[sort&compress]{natbib}
\bibpunct{(}{)}{;}{a}{}{,} 

\usepackage{amsthm, amsmath, amssymb, mathrsfs, multirow, url, subfigure}
\usepackage{graphicx} 
\usepackage{ifthen} 
\usepackage{amsfonts}
\usepackage[usenames]{color}
\usepackage{fullpage}
\usepackage{tikz}

\theoremstyle{plain} 
\newtheorem{theorem}{Theorem}
\newtheorem{corollary}{Corollary}
\newtheorem{proposition}{Proposition}

\theoremstyle{definition}

\theoremstyle{remark} 
\newtheorem{example}{Example}

\newtheorem{condition}{Condition}

\newcommand{\unif}{{\sf Unif}}
\newcommand{\nm}{{\sf N}}

\newcommand{\bet}{{\sf Beta}}

\newcommand{\XX}{\mathbb{X}}

\newcommand{\UU}{\mathbb{U}}

\newcommand{\TT}{\mathbb{T}}

\newcommand{\bigmid}{\; \Bigl\vert \;}

\newcommand{\model}{\mathcal{P}}

\newcommand{\convex}{\mathcal{Q}}

\newcommand{\pr}{\text{\sc pr}}

\title{Anytime valid and asymptotically efficient inference driven by predictive recursion}
\author{Vaidehi Dixit\footnote{Department of Statistics, University of Missouri, {\tt vdixit@missouri.edu}} \quad and \quad Ryan Martin\footnote{Department of Statistics, North Carolina State University, {\tt rgmarti3@ncsu.edu}}}
\date{\today}

\begin{document}

\maketitle

\begin{abstract}
Distinguishing two candidate models is a fundamental and practically important statistical problem.  Error rate control is crucial to the testing logic but, in complex nonparametric settings, can be difficult to achieve, especially when the stopping rule that determines the data collection process is not available.  This paper proposes an e-process construction based on the predictive recursion ($\pr$) algorithm originally designed to recursively fit nonparametric mixture models.  The resulting $\pr$e-process affords anytime valid inference and is asymptotically efficient in the sense that its growth rate is first-order optimal relative to $\pr$'s mixture model. 

\smallskip

\end{abstract}

\section{Introduction}
\label{S:intro}

A fundamental problem in statistics is that of distinguishing between two classes of candidate models based on observed data.  When the two classes of models are {\em simple}, i.e., there are just two distinct probability distributions being compared, then \citet{neymanpearson1933} settle the question on how to optimally distinguish the two based on the magnitude of the likelihood ratio.  Beyond the test's statistical properties, the {\em law of likelihood} \citep[e.g.,][]{edwards1992, royall.book, hacking.logic.book} offers principled justification for comparing the two models in this fashion.  In applications, however, it is rare for the relevant comparison to be between two simple hypotheses, and if either of the two hypotheses are {\em composite}, i.e., consisting of more than one probability distribution, then it is no longer clear how to operationalize the law of likelihood.  

Two common strategies are available to quantify the ``likelihood'' of a composite hypothesis: classical likelihood ratios, like those covered by Wilks's theorem \citep{wilks1938}, maximize the likelihood, whereas Bayes factors \citep[e.g.,][]{bernardo.smith.book, jeffreys1961} average the likelihood with respect to suitable priors.  In either case, judgments are made as above by considering the magnitude of the corresponding (maximum or average) likelihood ratio.  For various reasons, however, simply thresholding these likelihood ratios is not fully satisfactory: classical strategies to calibration are based on sampling distribution calculations that assume a specific data-generating process, and so the reliability statements lack robustness to the kind of departures from these assumptions common in practice, e.g., data-peeking.  And since neither of these likelihood ratios generally meet the conditions necessary (see below) to achieve a sort of universal calibration independent of the data-generating process, there is a need for new approaches.  

To meet this need, there has been a recent surge of interest in testing procedures that are both valid and efficient under optional stopping; see \citet{evalues.review} for a survey of these rapidly-moving developments.  The basic building blocks are {\em e-values}, which in some cases lead to {\em test martingales} \citep{shafer.vovk.martingale} and other kinds of {\em e-processes}.  These also have close connections to betting interpretations of probability \citep{shafer.betting} as well as {\em game-theoretic probability} \citep{shafer.vovk.book.2019}.  A practical challenge, however, is that there are so many e-values for data analysts to choose from in applications.  Two general and principled strategies for the construction of e-values (which are sometimes e-processes) are {\em universal inference} \citep{wasserman.universal} and {\em reverse information projection} \citep{grunwald.safetest}.  The former approach has different variants, but these can be computationally demanding and/or lacking in statistical efficiency.  The latter strategy is purely efficiency-motivated, and that makes it difficult to apply in nonparametric problems.  So, there is a need for an e-process construction that is flexible, conceptually and computationally simple, and is able to achieve the optimal first-order growth rate asymptotically.  The present paper offers such a construction. 

The familiar likelihood ratios fail to offer {\em anytime valid} inference in the sense of controlling errors independent of the stopping rule, etc.  As explained in Section~\ref{SS:evalues}, it turns out that e-processes themselves are likelihood ratios.  Since the null model is determined by the problem itself, an e-process construction reduces to specification of the non-null model likelihood.  Intuitively, one wants this non-null model likelihood to match the ``true likelihood'' as closely as possible. Of course, the data analyst does not know the true likelihood, so his/her choice of non-null likelihood must be sufficiently flexible to adapt to the information in the data concerning the underlying distribution.  Mixtures are flexible, so our proposal is to build this non-null likelihood using a nonparametric mixture model.  Fitting such a mixture model is computationally demanding using classical methods, but here we employ a fast, recursive algorithm called {\em predictive recursion}, or $\pr$ for short \citep[e.g.,][]{nqz, newton02}; see Section~\ref{SS:pr} for a review.  With this flexible and computationally efficient strategy to construct a non-null model likelihood, we define a corresponding e-process in Section~\ref{S:prevalues}, which we call the {\em $\pr$e-process}, that offers anytime valid and efficient statistical inference. 

Thanks to $\pr$'s naturally recursive model fit, the $\pr$e-process is easily upper-bounded by a genuine likelihood ratio, which is a test martingale under the null and hence has expected value upper-bounded by 1.  Then the anytime validity, e.g., hypothesis tests that are valid uniformly over stopping rules, is an immediate consequence; see Section~\ref{SS:valid}.  With anytime validity guaranteed, we turn our attention to efficiency in Section~\ref{SS:growth}.  There we establish that the $\pr$e-process asymptotically achieves the first-order optimal growth rate when the null is false.  


Two illustrations are presented in Section~\ref{S:examples}, where we show that the $\pr$e-process's empirical, finite-sample growth rate closely matches the theoretical growth rate advertised in Theorem~\ref{thm:efficient}.  Then, in Section~\ref{S:applications}, we consider two applications.  The first is a classical nonparametric problem of testing for log-concavity in the density, and we investigate the efficiency of our $\pr$e-process based test compared to the recently proposed tests in \citet{gangrade2023sequential} and \citet{dunn.etal.logconcave}.  The second application concerns testing for time-homogeneity in discrete-valued time series data.  Some concluding remarks are in Section~\ref{S:discuss} and further discussion and technical details can be found in the supplementary material. 



\section{Background}

\subsection{Notation}

The data consists of a sequence $X_1,X_2,\ldots$ of $\XX$-valued random variables, and we write $X^n = (X_1,\ldots,X_n) \in \XX^n$, for $n \geq 1$.  Let $\mathscr{A}$ denote the $\sigma$-algebra of measurable subsets of $\XX^\infty$ and write $\mathscr{A}_k = \sigma(X^k)$, for the filtration generated by $X^k$, $k \geq 1$.  An integer-valued random variable $N$ will be called a {\em stopping time} if $\{N \leq n\} \in \mathscr{A}_n$.  

Consider a collection $\model$ of candidate joint distributions for $X^\infty = (X_1, X_2, \ldots)$ on $\mathscr{A}$ and let $P=P^\infty$ denote a generic distribution in $\model$.  Write $\model_0$, a subset of $\model$, for the null hypothesis.  Throughout, we use upper-case letters for distributions/probability measures and the corresponding lower-case letters for the associated density (with respect to some common dominating measure).  For example, if $P_0$ is a member of $\model_0$, then $p_0$ is its corresponding density.

\subsection{E-processes}
\label{SS:evalues}

An {\em e-value}, $E$, relative to $P_0$, is a non-negative random variable with $P_0$-expected value (upper bounded by) 1.  
With batch data, say, $X$, and models that have densities, the non-negativity and expected value constraint implies that $x \mapsto q(x) := E(x) \, p_0(x)$ is a (sub-probability) density too; therefore, $E(x) = q(x)/p_0(x)$ is a likelihood ratio which can be interpreted as a measure of the evidence in data $x$ against $P_0$, relative to $Q$.  Consequently, if the observed value of $E$ is large, then one would be inclined to reject the hypothesis $P_0$.  A Bayes factor is an e-value relative to a single $P_0$; but likelihood ratios of the common form ``$\sup_p p(x) / p_0(x)$'' are not e-values.  

The situation described above is impractically simple.  Indeed, it is rare that $\model_0$ is a singleton and often data does not come in a batch.  {\em Test martingales} \citep{shafer.vovk.martingale} offer a solution to the non-batch data problem.  A test martingale is a non-negative martingale $(M_n)$, relative to $P_0$, adapted to the filtration $(\mathscr{A}_n)$ with $M_0 \equiv 1$.  Test martingales can also be expressed as likelihood ratios, i.e., if $Q$ is absolutely continuous with respect to $P_0$, then 
\begin{equation}
\label{eq:test.martingale}
M_n = \frac{q(X^n)}{p_0(X^n)} = \prod_{i=1}^n \frac{q(X_i \mid X^{i-1})}{p_0(X_i \mid X^{i-1})}, \quad n \geq 1. 
\end{equation}
So then the construction of a test martingale boils down to the choice of $Q$ for the numerator.  To handle composite $\model_0$ cases, one can construct a collection of test martingales $(M_n^{P_0})$ indexed by $n$ and by $P_0 \in \model_0$.  This also has a likelihood ratio form but, the choice of $Q$ could vary with $P_0$.  Now define an {\em e-process} $(E_n)$ as a sequence of non-negative random variables with $E_n \leq M_n^{P_0}$ for all $n$ and all $P_0 \in \model_0$.  An example of this would be $E_n = \inf_{P_0 \in \model_0} M_n^{P_0}$ and, if it happened that the $Q$ in \eqref{eq:test.martingale} does not depend on $P_0$, then this simplifies to 
\[ E_n = \frac{q(X^n)}{\sup_{P_0 \in \model_0} p_0(X^n)}, \quad n \geq 1. \] 
This is a common e-process construction \citep[e.g.,][Sec.~8]{wasserman.universal} but it is not the only one; see \citet{evalues.review} for more discussion on the various alternatives, in particular, the proposals in \citet{grunwald.safetest} and in \citet{waudbysmith.ramdas.bounded}.  We focus here on e-processes that, as above, only depend on a choice of $Q$ representing the alternative.

E-processes tend to be small under $\model_0$. In particular, {\em Ville's inequality} \citep[e.g.,][]{shafer.vovk.book.2019, howard.etal.2021} states that if $M_n$ is a test martingale for $\model_0$, then 
\[ \sup_{P_0 \in \model_0} P_0(M_N \geq \alpha^{-1}) \leq \alpha, \quad \text{for all $\alpha \in (0,1)$ and all stopping times $N$}. \]
Of course, the same is true for any e-process $E_n$ bounded by $M_n$.  The implications of this are far-reaching: it leads to statistical tests that are {\em anytime valid} in the sense that the reliability claims hold (basically) no matter how the investigator decides to conclude their study and perform the statistical analysis, thereby offering additional ``safety'' \citep[cf.,][]{grunwald.safetest} compared to procedures that only control the Type~I error rate for a particular sampling scheme.  For example, a test that rejects $\model_0$ based on data $X^n$ when $E_n \geq \alpha^{-1}$ controls the Type~I error rate {\em even if the investigator peeked at the data} when deciding whether to conclude the study at time $n$.

\subsection{Predictive recursion}
\label{SS:pr}

Here and in what follows, we assume $X_1,X_2,\ldots$ are iid.  Following the notation above, a general mixture model for the (common) marginal distribution defines the density function as 
\begin{equation}
\label{eq:mixture}
q^\Psi(x) = \int_\UU p_u(x) \, \Psi(du), \quad x \in \XX, 
\end{equation}
where $p_u$ is the density corresponding to a distribution $P_u$, with $\{P_u: u \in \UU\} \subseteq \model$, and $\Psi$ is a mixing distribution defined on $\UU$. Such models are incredibly flexible, so they are often used in contexts where the mixture structure itself is not directly relevant, e.g., in density estimation applications.  It is common to think of $x \mapsto p_u(x)$ as a parametric kernel, like Gaussian, but that is not necessary here; indeed, $\UU$ could just be a generic indexing of the entire model $\model$. But we should emphasize that mixtures of simple parametric kernels can be incredibly flexible. For example, the set of Gaussian location--scale mixtures is dense with respect to total variation distance in the space of smooth densities in Euclidean space \citep[e.g.][Theorem~33.1]{dasgupta}.  

For fitting the mixture model \eqref{eq:mixture} to the observed data $X^n = (X_1,\ldots,X_n)$, one common strategy is nonparametric maximum likelihood \citep[e.g.,][]{laird, lindsay1995}. Another common strategy is to introduce a prior distribution for $\Psi$, e.g., a Dirichlet process prior \citep[e.g.,][]{ferguson1973, ghosal2010, lo1984}, and carry out a nonparametric Bayesian analysis.  There are advantages to sticking with existing approaches, the complexity of these models creates computational challenges.  In particular, since tailored Monte Carlo methods are required, one cannot capitalize on the Bayesian coherent updating property---``today's posterior is tomorrow's prior''---when data are processed sequentially.  Maximum likelihood estimation faces similar challenges. A novel alternative, developed by Michael Newton and collaborators in the late 1990s, is a fast, recursive algorithm called {\em predictive recursion}, or $\pr$ for short \citep[e.g.,][]{newton02}.  Applications in large-scale inference settings include \citet{taonewton1999}, \citet{newton-ecoli}, \citet{mt-test}, \citet{scott.fdrsmooth}, and \citet{woody.scott.post}.  

Here we provide a quick review of $\pr$ and its properties; for more details, see \citet{prjkg}.  Start with a weight sequence $(w_i: i \geq 1) \subset (0,1)$ that satisfies
\begin{equation}
\label{eq:weights.main}
\sum_{i=1}^\infty w_i = \infty \quad \text{and} \quad \sum_{i=1}^\infty w_i^2 < \infty. 
\end{equation}
Following \citet{mt-rate} and the discussion at the end of Section~\ref{SS:growth} below, we take $w_i = (i+1)^{-0.67}$ for all of our examples.  Next, take an initial guess $\hat\Psi_0$ of $\Psi$ supported on the index space $\UU$.  Then $\pr$ updates the initial guess along the data sequence as follows: 
\begin{equation}
\label{eq:pr}
\hat\Psi_i(du) = (1-w_i) \, \hat\Psi_{i-1}(du) + w_i \, \frac{p_u(X_i) \, \hat\Psi_{i-1}(du)}{\int_\UU p_v(X_i) \, \hat\Psi_{i-1}(dv)}, \quad u \in \UU, \quad i \geq 1. 
\end{equation}
$\pr$ is recursive, so one only needs $\hat\Psi_n$ and the new data point $X_{n+1}$ to get the new $\hat\Psi_{n+1}$.  Moreover, this leads naturally to a plug-in estimator of the mixture density 
\[ \hat q_{x^n}(x) := q^{\hat\Psi_n}(x) = \int_\UU p_u(x) \, \hat\Psi_n(du), \quad x \in \XX. \]
Large-sample properties of $\pr$ have been explored in \citet{ghoshtokdar}, \citet{martinghosh}, \citet{tmg}, \citet{mt-rate, mt-prml}, and \citet{dixit.martin.revisiting}.  The property most relevant to our efforts here will be explained in Section~\ref{S:prevalues}.  

The above display defines a $\pr$-based predictive density for $X_{n+1}$, given $X^n$. 
It also produces a joint marginal density for $X^n$---``marginal'' in the sense that the mixing distribution $\Psi$ has been integrated out---and it has a multiplicative form, \`a la \citet[][Eq.~10.10]{wald.sequential}, \citet{dawid1984}, and \citet{rissanen1984}.  That is, the joint marginal density $\hat q^\text{\sc pr}$ satisfies 
\begin{equation}
\label{eq:mult}
\hat q^\text{\sc pr}(x^n) = \hat q_{x^{n-1}}(x_n) \, \hat q^\text{\sc pr}(x^{n-1}) = \prod_{i=1}^n \hat q_{x^{i-1}}(x_i), \quad n \geq 1, \quad x^n \in \XX^n. 
\end{equation}
It is $\pr$'s flexibility and the multiplicative form of its joint marginal density that makes it especially suitable for anytime valid nonparametric inference.  

The simple $\pr$ algorithm is computationally efficient in the sense that the update $\hat\Psi_{n} \to \hat\Psi_{n+1}$ is an $O(1)$ operation in $n$.  But the integration in \eqref{eq:pr} must be done numerically and requires care.  
If the dimension of $\UU$ is relatively low, then integration can be handled easily and accurately using quadrature.  Monte Carlo would be a natural alternative, but direct sampling from $\hat\Psi_{n}$ is a challenge, so we developed a ``$\pr$ticle filter'' strategy \citep{prticle} that can easily accommodate mixtures over $\UU$ of moderate dimension, as in Example~\ref{ex:parametric} below and in the supplementary material.  Extensions to high-dimensional $\UU$ is a focus of ongoing research. 

The computations in \eqref{eq:pr} are carried out pointwise on a grid spanning $\UU$, which effectively requires $\UU$ to be fixed {\em a priori} and bounded.  In our experience, particularly with batch data, one can safely fix $\UU$ as a large compact that contains those $u$'s even remotely compatible with the observed data.  With sequential data, one could first use a block of data to inform the choice of $\UU$, then go back and process the full sequence with that choice of $\UU$.  More formally, \citet[][Sec.~5]{mt-rate} present a {\em generalized predictive recursion} algorithm that features an adaptive support, but we will not consider this here.

\section{PRe-processes}
\label{S:prevalues}

\subsection{Construction}

Consider null hypothesis $H_0: P \in \model_0$.  The proposal here is to construct a suitable marginal likelihood under $\model_0^c$ by mixing over a specified class $\{P_u: u \in \UU\}$ of distributions.  
Instead of a computationally demanding nonparametric maximum likelihood or Bayes fit, here we make use of the efficient $\pr$ algorithm reviewed in Section~\ref{SS:pr} above.  

Let $\Psi$ be a mixing probability distribution supported on the specified index set $\UU$ and consider a basic mixture model as in \eqref{eq:mixture}.  This model can be fit to data $X^n$ using $\pr$, and the output most relevant to us here is the joint marginal density for $X^n$ that is produced as a by-product, namely, $\hat q^\text{\sc pr}(X^n)$ as given in \eqref{eq:mult}.  Recall the multiplicative form of $\hat q^\text{\sc pr}(X^n)$ in \eqref{eq:mult}, a driving force behind the general formulation in \citet[][Sec.~8]{wasserman.universal}.  Moreover, the $\pr$ update from $\hat q^\text{\sc pr}(X^{n-1})$ to $\hat q^\text{\sc pr}(X^n)$ is an $O(1)$ computation, compared to the $O(n)$ computation expected with ``non-anticipatory'' maximum likelihood estimation \citep[e.g.,][]{gangrade2023sequential}. 

With this, we define the following $\pr$-driven e-process, i.e., {\em $\pr$e-process}, 
\begin{equation}
\label{eq:prevalue}
E_n^\text{\sc pr} = E^\text{\sc pr}(X^n; \model_0) = \frac{\hat q^\text{\sc pr}(X^n)}{\sup_{p \in \model_0} p(X^n)}, \quad n \geq 1. 
\end{equation}
Intuitively, since the likelihood tends to favor the true hypothesis, if $H_0$ is true (resp.~false), then $E_n^\text{\sc pr}$ ought to be small (resp.~large).  This intuition suggests a test that rejects $H_0$ if and only if $E_n^\text{\sc pr}$ is large, and we justify this intuition in Section~\ref{SS:valid} below.  

Our proposal can be compared to two of the now-standard e-process constructions reviewed in \citet{evalues.review}.  Specifically, our proposal is, on the one hand, like the basic {\em method of mixtures} \citep[e.g.,][]{wald1945, darling.robbins.1968} in that it forms a marginal likelihood under the alternative via a suitable mixture model, which we think is intuitively appealing.  To achieve both the flexibility and the appealing intuition, the mixture model needs to be nonparametric, which would pose computational challenges for traditional likelihood-based methods.  But $\pr$ is specifically designed to simply and efficiently handle this challenge.  So, the $\pr$e-process is like (the no data-splitting variants of) universal inference \citep[][Sec.~8]{wasserman.universal}, just with a specific focus on flexible mixture model alternatives with a fast, recursive updating scheme.  

Our proposal can also be compared to that in \citet{grunwald.safetest}.  While our choice to use ``$\sup_{p \in \model_0} p(X^n)$'' in the denominator of \eqref{eq:prevalue} is a natural one, it is not the only option.  Gr\"unwald et al.~propose a strategy based on the {\em reverse information projection} ({\sc rip}), as in \citet{libarron}, which amounts to replacing ``$\sup_{p \in \model_0} p(X^n)$'' in the denominator with the likelihood at a fixed---but strategically chosen---density, say, $p_0^\text{\sc rip}$ in the convex hull, $\text{co}(\model_0)$, of $\model_0$.  In our context, they might propose to make inference based on the ratio 
\[ E_n^\text{\sc pr+rip} = \hat q^\text{\sc pr}(X^n) \, / \, p_0^\text{\sc rip}(X^n), \]
where $P_0^\text{\sc rip}$ satisfies $K(\widehat Q^\text{\sc pr}, P_0^\text{\sc rip}) = \inf_{P_0 \in \text{co}(\model_0)} K(\widehat Q^\text{\sc pr}, P_0)$.  
This strategy has powerful motivation and nice properties concerning maximal growth rate.  However, the complexity of the applications we have in mind, along with $\pr$'s non-trivial learning process, raise some difficult questions: first, how to compute $P_0^\text{\sc rip}$ and, second, does $E_n^\text{\sc pr+rip}$ define a proper e-process?  We leave these questions for future investigation.

\subsection{Validity}
\label{SS:valid}

Of course, we cannot refer to the quantity defined in \eqref{eq:prevalue} as an ``e-process'' without showing that it satisfies the required properties.  Theorem~\ref{thm:valid} establishes the basic e-process property from which all the relevant statistical properties follow. 

\begin{theorem}
\label{thm:valid}
Consider a model $\model$ for the iid data sequence $X_1,X_2,\ldots$.  For a model $\model_0 \subset \model$ to be tested, the $\pr$e-process defined in \eqref{eq:prevalue} is an e-process. 
\end{theorem}

\begin{proof}
For any fixed $n$, since the supremum over $\model_0$ appears in the denominator of \eqref{eq:prevalue}, the following inequality is immediate:
\[ E_n^\text{\sc pr} = E^\text{\sc pr}(X^n; \model_0) \leq E^\text{\sc pr}(X^n; \{p_0\}) =  \hat q^\text{\sc pr}(X^n) \, / \, p_0(X^n), \quad \text{for all $P_0 \in \model_0$}. \]
The upper bound is a collection of test martingales indexed by $P_0 \in \model_0$ and, therefore, $(E_n^\text{\sc pr})$ is an e-process under $\model_0$.  
\end{proof}

From here, we can immediately deduce several directly interpretable statistical results.  In particular, suitably-defined $\pr$e-process-based testing procedures are anytime valid.  

\begin{corollary}
\label{cor:test}
For a desired significance level $\alpha \in [0,1]$, let $T_\alpha(X^n)$ be the test that rejects $H_0: P \in \model_0$ if and only if $E^\text{\sc pr}(X^n; \model_0) \geq \alpha^{-1}$.  This test controls the frequentist Type~I error at the designated level, i.e., $\sup_{P_0 \in \model_0} P_0\{ T_\alpha(X^N) \text{ rejects} \} \leq \alpha$ for any stopping rule $N$. 
\end{corollary}


If one has a valid test, then of course it can be inverted to construct valid confidence sets.  Beyond these familiar statistical procedures, one can  leverage the $\pr$e-process for the purpose of broader uncertainty quantification, as discussed in, e.g., \citet{martin.partial2, martin.basu} and \citet{grunwald.beyond, grunwald.epost}.  For further details on these, see the supplementary material.

\subsection{Asymptotic growth rate}
\label{SS:growth}

Theorem~\ref{thm:valid} establishes that $\pr$e-process-based tests, etc.~are anytime valid.  For the $\pr$e-process procedure to be {\em efficient}, we want the e-value to be large under the alternative, at least asymptotically, so that we will correctly reject false null hypotheses.  Our goal here is to show that the $\pr$e-process is asymptotically optimal in the sense of \citet[][Ch.~11]{cover.thomas.book}, i.e., its asymptotic growth rate is optimal to first order in the exponent.  Towards this, consider the case where $P^\star \not\in \model_0$ determines the true distribution of the data $X^\infty$. So far, we have been implicitly assuming that mixtures of the kernels $\{P_u: u \in \UU\}$ are sufficiently flexible to warrant consideration as a ``non-null model,'' i.e., one is willing to use such a model for nonparametric estimation of the true distribution; see Section~\ref{SS:pr}.  If $\convex$ denotes the convex hull of $\{P_u: u \in \UU\}$, the set of all mixtures of these kernels, then a more explicit statement of the aforementioned implicit assumption is that $\convex$ is an acceptable non-null model.  In other words, if one believes that $P^\star \not\in\model_0$, then he/she would also believe that $P^\star$ is ``closer'' to $\convex$ than to $\model_0$.

The main result in \citet{mt-prml} states that, under certain regularity conditions (see the supplementary material), with $P^\star$-probability~1, 
\begin{equation}
\label{eq:pr.limit.main}
n^{-1} \log \hat q^\text{\sc pr}(X^n) = n^{-1} \log p^\star(X^n) - K(P^\star, \convex) + o(1), \quad n \to \infty, 
\end{equation}
where $K(P^\star, Q)$ is the Kullback--Leibler divergence of $Q$ from $P^\star$ and
\begin{equation}
\label{eq:inf}
K(P^\star, \convex) = \inf_{Q \in \convex} K(P^\star, Q),
\end{equation}
with $\convex = \text{co}(\{P_u: u \in \UU\})$ the set of mixtures.  For the null $\model_0$, let $\kappa^\star(\model_0)$ be such that 
\begin{equation}
\label{eq:mle.limit.main}
\liminf_{n \to \infty} n^{-1} \log\{ p^\star(X^n) / \hat p_0(X^n)\} \geq \kappa^\star(\model_0), \quad \text{with $P^\star$-probability 1}, 
\end{equation}
where $\hat p_0(X^n) = \sup_{P_0 \in \model_0} p_0(X^n)$ is the maximum likelihood estimator under the null $\model_0$.  From the definition of $\hat p_0$, it follows that $\kappa^\star(\model_0) \leq K(P^\star, \model_0) := \inf_{P_0 \in \model_0} K(P^\star, P_0)$, and often $\kappa^\star(\model_0)$ will equal this minimal Kullback--Leibler number.  For example, if $\model_0 = \{P_0\}$ is a singleton and $K(P^\star, P_0) < \infty$, then the law of large numbers gives \eqref{eq:mle.limit.main} with ``$\lim$,'' 
``$=$,'' and $\kappa^\star(\model_0) = K(P^\star, P_0)$.  More generally, suppose there exists $P_0^\dagger \in \model_0$ with $K(P^\star, P_0^\dagger) = K(P^\star, \model_0)$; see, e.g., \citet{patilea2001} and \citet{kleijn}.  Then 
\[ n^{-1} \log\{ p^\star(X^n) / \hat p_0(X^n) \} = n^{-1} \log\{ p^\star(X^n) / p_0^\dagger(X^n) \} + n^{-1} \log\{ p_0^\dagger(X^n) / \hat p_0(X^n) \}. \]
The first term on the right-hand side converges to $K(P^\star, \model_0)$ with $P^\star$-probability 1, so, the bound in \eqref{eq:mle.limit.main} is determined by the second term on the right-hand side.  That term vanishes under relatively mild conditions on $\model_0$, e.g., when the maximum likelihood estimator is consistent for $P_0^\dagger$ (relative to the true distribution $P^\star$).  In these typical cases, \eqref{eq:mle.limit.main} holds with ``$\lim$,'' ``$=$,'' and $\kappa^\star(\model) = K(P^\star, \model_0)$.  But it is possible that $\kappa^\star(\model_0) < K(P^\star, \model_0)$, e.g., in cases where consistency fails; see the supplementary material for an example. 

Putting all this together, with $P^\star$-probability~1, the $\pr$e-process \eqref{eq:prevalue} satisfies 
\begin{align*}
n^{-1} \log E_n^\text{\sc pr} & = n^{-1} \log\{ \hat q^\text{\sc pr}(X^n) / \hat p_{0}(X^n)\} \\
& = n^{-1} \log\{ p^\star(X^n) / \hat p_{0}(X^n) \} + n^{-1} \log\{ \hat q^\text{\sc pr}(X^n) / p^\star(X^n)\} \\
& \geq \kappa^\star(\model_0) - K(P^\star, \convex) + o(1).
\end{align*}
More details about \eqref{eq:pr.limit.main}, \eqref{eq:mle.limit.main}, and the proof are given in the supplementary material.  

\begin{theorem}
\label{thm:efficient}
Consider $X_1,X_2,\ldots$ iid with distribution $P^\star$.  If \eqref{eq:pr.limit.main} and \eqref{eq:mle.limit.main} hold, then with $P^\star$-probability~1, the $\pr$e-process has asymptotic first-order growth rate 
\begin{equation}
\label{eq:growth.rate}
\Delta(P^\star; \model_0, \convex) = \kappa^\star(\model_0) - K(P^\star, \convex). 
\end{equation}
That is, as $n \to \infty$, the $\pr$e-process satisfies 
\[ \log E_n^\text{\sc pr} \geq n \, \Delta(P^\star; \model_0, \convex) + o(n), \quad \text{with $P^\star$-probability 1.} \]
\end{theorem}

A few remarks about Theorem~\ref{thm:efficient} are in order.  We focus our attention here on the typical case where $\kappa^\star(\model_0) = K(P^\star, \model_0)$.  First, the non-null case where $K(P^\star, \model_0) > K(P^\star, \convex)$ is of primary interest. This, again, is because $\convex$ is judged to be an acceptable nonparametric model, so surely a non-null $P^\star$ is closer to $\convex$ than to $\model_0$. We claim that the growth rate $\Delta(P^\star; \model_0, \convex)$ is ``optimal'' which deserves explanation.  In the simple-versus-simple case, with $\model_0 = \{P_0\}$ and $\convex = \{P_1\}$, ``the likelihood ratio is growth rate optimal'' \citep{evalues.review} and its growth rate $K(P_1, P_0)$ agrees with \eqref{eq:growth.rate}. 
More generally, given a commitment to maximum $\model_0$-likelihood in the e-process denominator, the user's only input in the e-process construction is the choice of ``$\hat q(X^n)$'' in the numerator.  If this $\hat q$ is constrained to $\convex$, then there is no choice of numerator that can achieve a larger first-order asymptotic growth rate than that in \eqref{eq:growth.rate}.  This can also be compared to the growth rates in \citet{grunwald.safetest}.  Indeed, if the true $P^\star$ was {\em known} and used in their e-value's numerator, then their solution is, by definition, growth rate optimal and its first-order growth rate agrees with that in Theorem~\ref{thm:efficient}.  Since the growth rate cannot be better when the true $P^\star$ is unknown, our $\pr$e-process must be asymptotically optimal in the above sense too.  

Second, suppose $K(P^\star, \model_0) < K(P^\star, \convex)$, i.e., $\model_0$ is ``more true'' then $\convex$; a special case is when $K(P^\star, \model_0) = 0$ and hence $\model_0$ is ``true.'' Then we want the $\pr$e-process to be vanishing with $n$ and, in the not-uncommon case where \eqref{eq:mle.limit.main} holds with ``$\lim$'' and ``$=$,'' Theorem~\ref{thm:efficient} establishes this: if $K(P^\star, \model_0) < K(P^\star, \convex)$, then the $\pr$e-process vanishes as $n \to \infty$; see, also, \citet[][Sec.~8.2]{ramdas.nnm}.  In the edge case where $K(P^\star, \model_0)=K(P^\star, \convex)$, so $\model_0$ and $\convex$ are ``equally true'' and cannot be distinguished, we do not specifically expect $E_n^\text{\sc pr}$ to grow or to vanish, so Theorem~\ref{thm:efficient}'s ambiguity about the behavior of $E_n^\text{\sc pr}$ in such cases accurately reflects this.  

Next, how small is the $o(n)$ term in Theorem~\ref{thm:efficient}'s lower bound?  This is relevant because it describes the asymptotic gap between the proposed $\pr$e-process and an oracle e-process that uses the true $p^\star$ in the numerator.  Specifically, the oracle e-process is given by 
\[ E_n^\text{\sc or} = \frac{p^\star(X^n)}{\sup_{P_0 \in \model_0} p_0(X^n)}, \]
and, asymptotically, the difference between the two log-e-processes, $\log E_n^\text{\sc or} - \log E_n^\text{\sc pr}$, is the ``$o(n)$'' term in our lower bound.  Moreover, that same log-difference is 
\[ \log E_n^\text{\sc or} - \log E_n^\text{\sc pr} = \log \frac{p^\star(X^n)}{\hat q^\text{\sc pr}(X^n)} = \sum_{i=1}^n \log \frac{p^\star(X_i)}{\hat q_{X^{i-1}}(X_i)}, \]
where $\hat q_{X^{i-1}}$ is $\pr$'s one-step-ahead predictive density as in \eqref{eq:mult}.  \citet{mt-prml} showed that the right-hand side of the above display grows, asymptotically, like $n K(p^\star, \hat q_{X^n})$. 
Furthermore, \citet{mt-rate} proved the following: in addition to the conditions presented in the supplementary material, if $P^\star$ is in the interior of $\convex$ and if $\pr$'s weight sequence $(w_i)$ is such that $w_n \asymp n^{-\gamma}$ for $\gamma \in (2/3,1]$, then $n^{1-\gamma} K(p^\star, \hat q_{X^n}) \to 0$ with $P^\star$-probability 1. One immediate take-away is that, apparently, $\gamma \approx 2/3$ gives the best rates for $\pr$.  Those authors argue that their rate is ``worst-case'' corresponding to $P^\star$ arbitrarily close to the boundary of $\convex$, 
which is most difficult for $\pr$ to learn.  When $P^\star$ is away from the boundary of $\convex$, however, $\pr$'s rate tends to be faster than Martin \& Tokdar's conservative upper bound.  In fact, our empirical work suggest the following conjecture: $n^{1-\gamma/2} K(p^\star, \hat q_{X^n}) \to 0$; see Section~\ref{S:discuss}.  If true, then this would be near-optimal in certain cases.  For instance, if $p^\star$ is a monotone decreasing density, i.e., a scale mixture of uniform kernels \citep[e.g.,][]{williamson1956}, then $\pr$'s rate with $\gamma \approx 2/3$ would virtually agree with the well-known minimax optimal rate of $n^{-2/3}$ in Kullback--Leibler divergence.  (There is another new sense in which $\pr$'s rate is actually faster than what Martin \& Tokdar's result suggests, which we present in Section~4 of the supplementary material.)  Returning to the original question, the aforementioned theory implies that, with $\gamma \approx 2/3$, the $\pr$e-process's $o(n)$ error term is conservatively like $n^{2/3}$.  Our conjecture, however, says that the error is more like $n^{1/3}$, and the empirical results in Section~5 of the supplementary material support this.  Better approximations of the oracle e-process are possible in some cases, e.g., using nonparametric Bayes predictive densities in the e-process numerator, but not with $\pr$'s computational efficiency---and the corresponding ``$o(n)$'' term would still generally be of polynomial order in $n$.

\section{Illustrations}
\label{S:examples}


Here we consider two illustrations of the procedure described in Section~\ref{S:prevalues}: testing for monotonicity and testing a parametric model. Our focus here is on comparing the empirical growth rate of the $\pr$e-process to that predicted by Theorem~\ref{thm:efficient}.  So, in what follows, the data $X_1,X_2,\ldots$ will be generated from a distribution $P^\star$ that does not belong to $\model_0$.

\begin{figure}[t]
\centering
\subfigure[Example~\ref{ex:monotone}: testing monotonicity]
{\includegraphics[width = 0.45\linewidth]{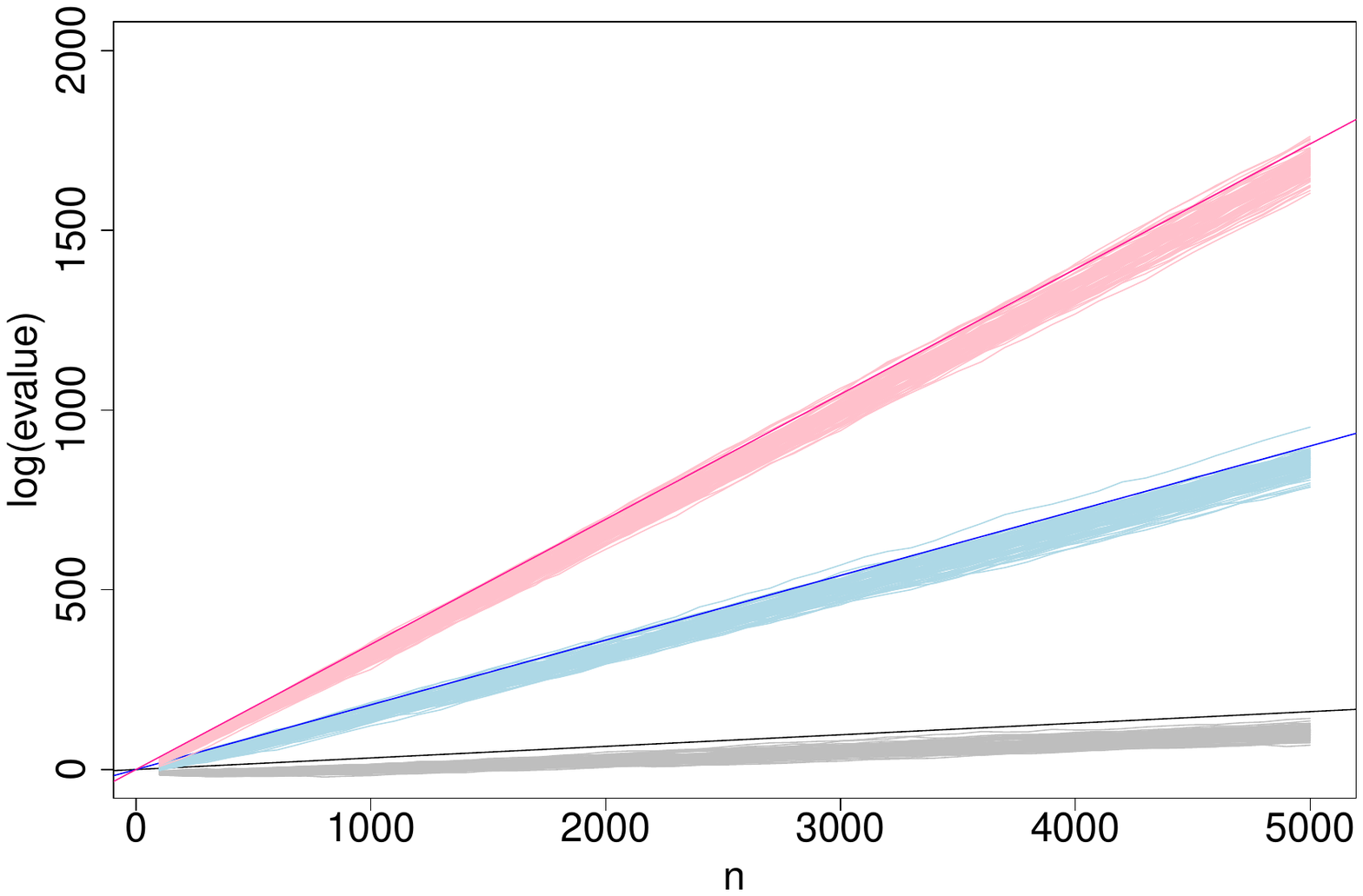}
\label{fig:monotone}}
\subfigure[Example~\ref{ex:parametric}: testing a parametric null]
{\includegraphics[width = 0.45\linewidth]{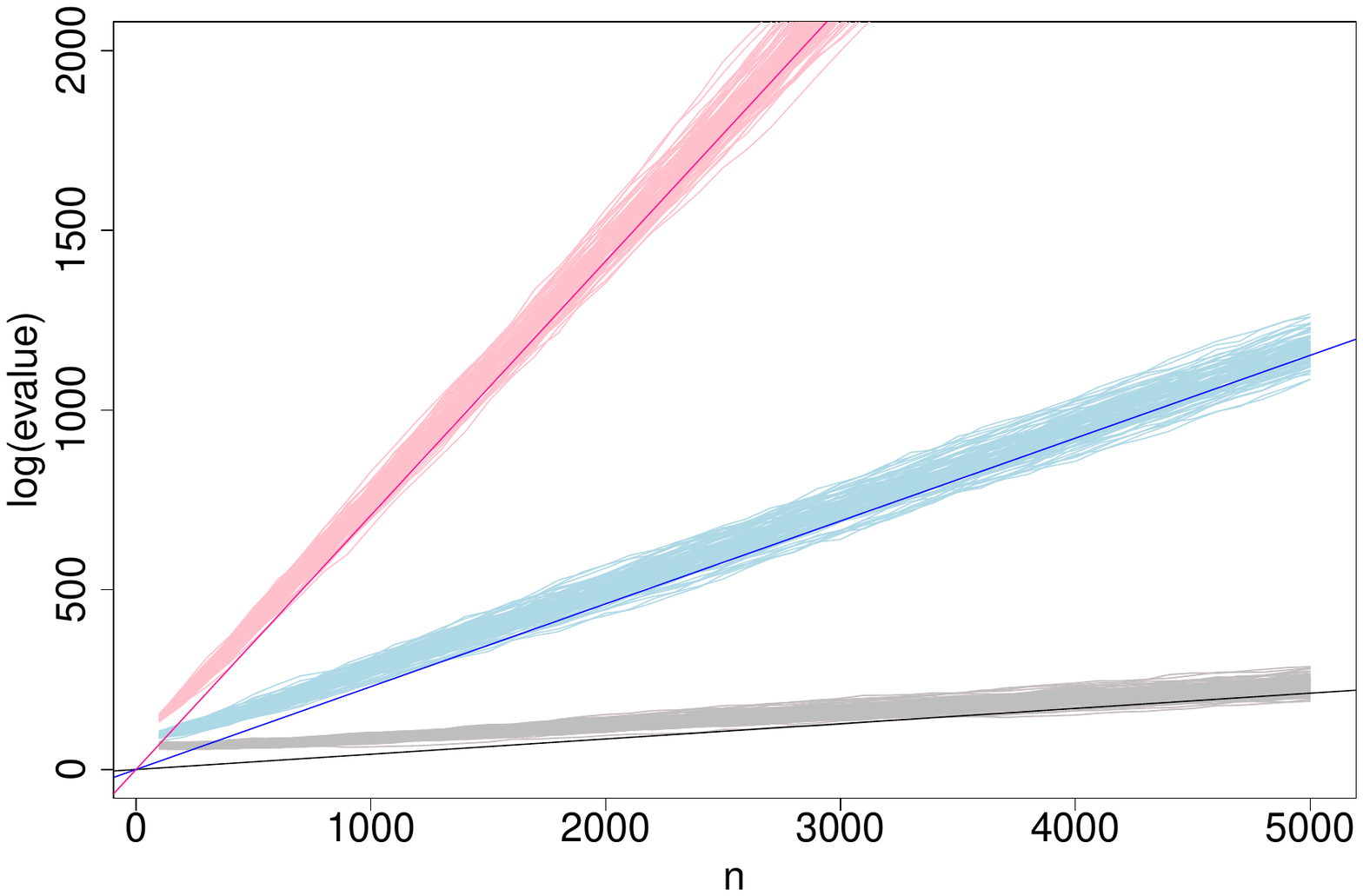}
\label{fig:vmf}}
\caption{
Plots of $\log E_n^\text{\sc pr}$ versus $n$ for $100$ data sets of size $n=5000$ under three settings of the true density in the alternative. Each case shows a line with slope based on the theoretical growth rate in Theorem~\ref{thm:efficient}.  In both panels, the gray, blue, and pink lines correspond to cases where the data are generated from a distribution that is ``close,'' ``less close,'' and ``farthest'' from the null, respectively. Specific details about the settings are given in the text.
}
\label{fig:evalue}
\end{figure}

\begin{example}
\label{ex:monotone}
Monotone densities on the positive half-line are common in biomedical, engineering, and astronomy applications.  
Our goal is to test the null hypothesis that the underlying density is monotone.  
For the denominator of our $\pr$e-process, we use R package {\tt REBayes} \citep{rebayes} to find the the nonparametric maximum likelihood estimator, i.e., the Grenander estimator \citep[e.g.,][]{grenander}.  For the numerator, we use a mixture model with a gamma kernel, i.e., $p_u$ is a gamma density with $u$ the shape and rate parameter pair.  Then we fit a mixture model over $\UU = [1,15] \times [10^{-5} ,5]$ with the initial guess $\Psi_0$ the uniform distribution over $\UU$. For our experiments, the true distribution $P^\star$ is a gamma distribution with unit rate parameter and varying shape parameter.  The idea is that, if the shape parameter is 1, then $P^\star$ would be exponential which is monotone; so as the shape parameter varies from 2, to 5, and to 10, the density becomes ``less monotone.''  We generate 100 data sets from $P^{\star}$ and calculate the $\pr$e-process $E_n^\text{\sc pr}$ at the increments $n = 100, 200, \ldots, 5000$. 
A plot of the $\log E_n^\text{\sc pr}$ versus $n$ is displayed in Figure~\ref{fig:monotone}, along with a reference line having slope $K(P^{\star}, \model_0)$. First note that, as expected, the growth rate increases as the true density gets ``less monotone'' and, second, that the $\pr$e-process closely follows the theoretical growth rate across all three scenarios.
\end{example}

\begin{example}
\label{ex:parametric}
Testing for clustering of directional data is an important practical problem \citep[e.g.,][]{mardia.jupp.book}. Here we focus on testing for a single von Mises--Fisher model $\model_0$ versus a mixture thereof.  This problem is also challenging since the null lies on the boundary of the alternative \citep[e.g.,][]{mcvinish.rousseau.mengersen.2009, tokdar.chakrabarti.ghosh.2010}.  

Computation of the $\pr$e-process's denominator is straightforward, since closed-form expressions are available for the maximum likelihood estimators. The von Mises--Fisher distribution is generally parameterized with a mean parameter $\mu$ and concentration parameter $\kappa$, where $\|\mu \| \leq 1$ and $\kappa \geq 0$. So for the numerator we take the mixing distribution variable as $u = (\mu_\theta, \mu_\phi, \kappa)$ where 
$(\mu_\theta, \mu_\phi) \in (0,\pi)\times(0,2\pi)$ represent the spherical coordinates of $\mu$. Note that because of the multivariate nature of the problem we need to implement the $\pr$ticle filter machinery in \citet{prticle} to calculate the normalizing constants. For this we generate particles of size $t=10,000$ from $\unif(0, \pi)$, $\unif(0, 2\pi)$ and $\unif(0.2, 10)$ respectively for $\mu_\theta, \mu_\phi$ and $\kappa$. 
For our simulation, we generate data from a bimodal mixing distribution $\Psi^\star$ which fixes $\kappa = 10$ and distributes $(\mu_\theta, \mu_\phi)$ over $(0,\pi)\times(0,2\pi)$ according to a mixture of truncated bivariate normals,
\begin{equation}
\label{eq:mixvmf}
   (\mu_\theta, \mu_\phi) \sim 0.5 \, \mathsf{trN}_2(\eta_1, \Sigma) + 0.5 \,  \mathsf{trN}_2(\eta_2, \Sigma).
\end{equation}
In \eqref{eq:mixvmf}, the covariance matrix $\Sigma$ is a diagonal matrix with the diagonal vector as 
$\{(\pi/12)^2, (\pi/6)^2\}$, and the mean vectors are $\eta_1 = (\pi/4, \pi/2)^\top$ and $\eta_2$ varying, 
\[\eta_2 \in \{(3\pi/8, 3\pi/4)^\top, (\pi/2, \pi)^\top, (3\pi/4, 3\pi/2)^\top\}\].  
The distance between the two modes $(\eta_1, \eta_2)$ acts as a measure of the ``degree of clustering,'' with large distances corresponding to going {\em further away} from the null. We generate 100 data sets under each configuration and plot $\log E_n^\text{\sc pr}$ versus $n$ at increments $n=100, 200, \ldots, 5000$ in Figure~\ref{fig:vmf}, along with a reference line of slope $K(P^{\star}, \model_0)$. As expected, the slope increases in the ``degree of clustering'' and the log-$\pr$e-process paths closely follow this trend.  
\end{example}

\section{Applications}
\label{S:applications}

\subsection{Testing log-concavity of a density}

Examples of log-concave densities on the real line include the Gaussian, logistic, Laplace, and others.  Efficient numerical methods have been developed \citep[e.g.,][]{dumbgen2011logcondens, cule.etal.2010} to find the maximum likelihood estimator of a log-concave density, and its asymptotic properties are studied in, e.g., \citet{doss.wellner.2016}.  As for testing log-concavity, e-process-based tests have been proposed recently in \citet{gangrade2023sequential} and \citet{dunn.etal.logconcave}.  Here, we compare the $\pr$e-process's growth rate to that of the aforementioned e-processes.  

For the $\pr$e-process denominator, we get the log-concave maximum likelihood estimator using the {\tt logConDens} function in the R package 
{\tt logcondens} \citep{dumbgen2011logcondens}. For the numerator, we consider a mixture model with Gaussian kernel $p_u$, where $u$ is mean and standard deviation pair, and mixing distribution supported on $\UU = [-10, 20] \times [0.01, 3]$. An R package for $\pr$e-process testing of log-concavity can be found at {\tt https://github.com/vdixit005/PReprocess}. For Dunn et al.'s e-process, $E_n^\text{\sc ui}$, we use the code at {\tt https://github.com/RobinMDunn/LogConcaveUniv} with its default settings. For the e-process proposed in \citet{gangrade2023sequential}, $E_n^\text{\sc ulr}$, we follow their recommendation and use an iteratively fit Gaussian kernel density estimator.  

\begin{figure}[t]
\centering
\subfigure[$\mu=6$]
{\includegraphics[width = 0.315\linewidth]{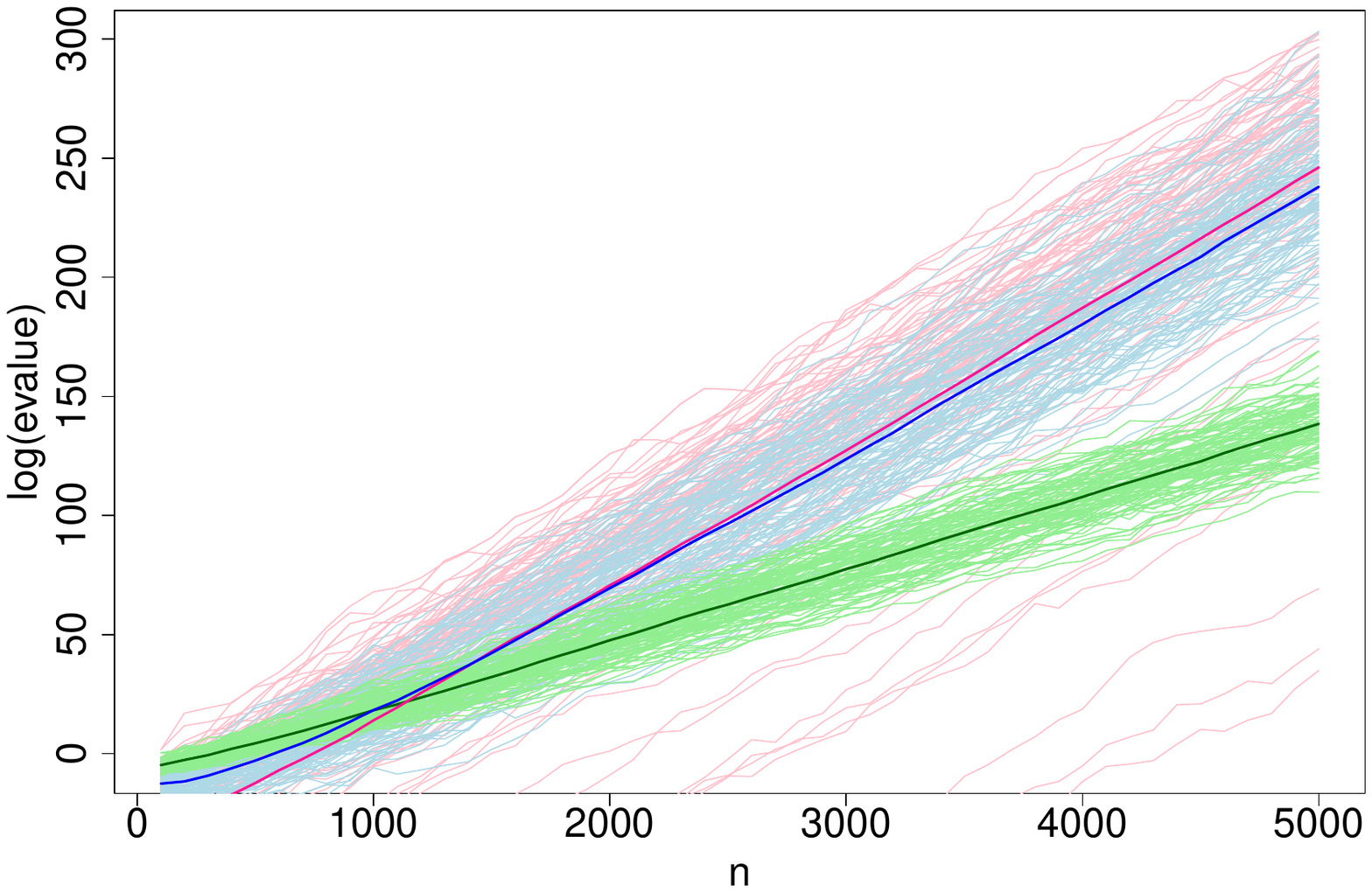}
\label{fig:logconcave1}}
\subfigure[$\mu=10$]
{\includegraphics[width = 0.315\linewidth]{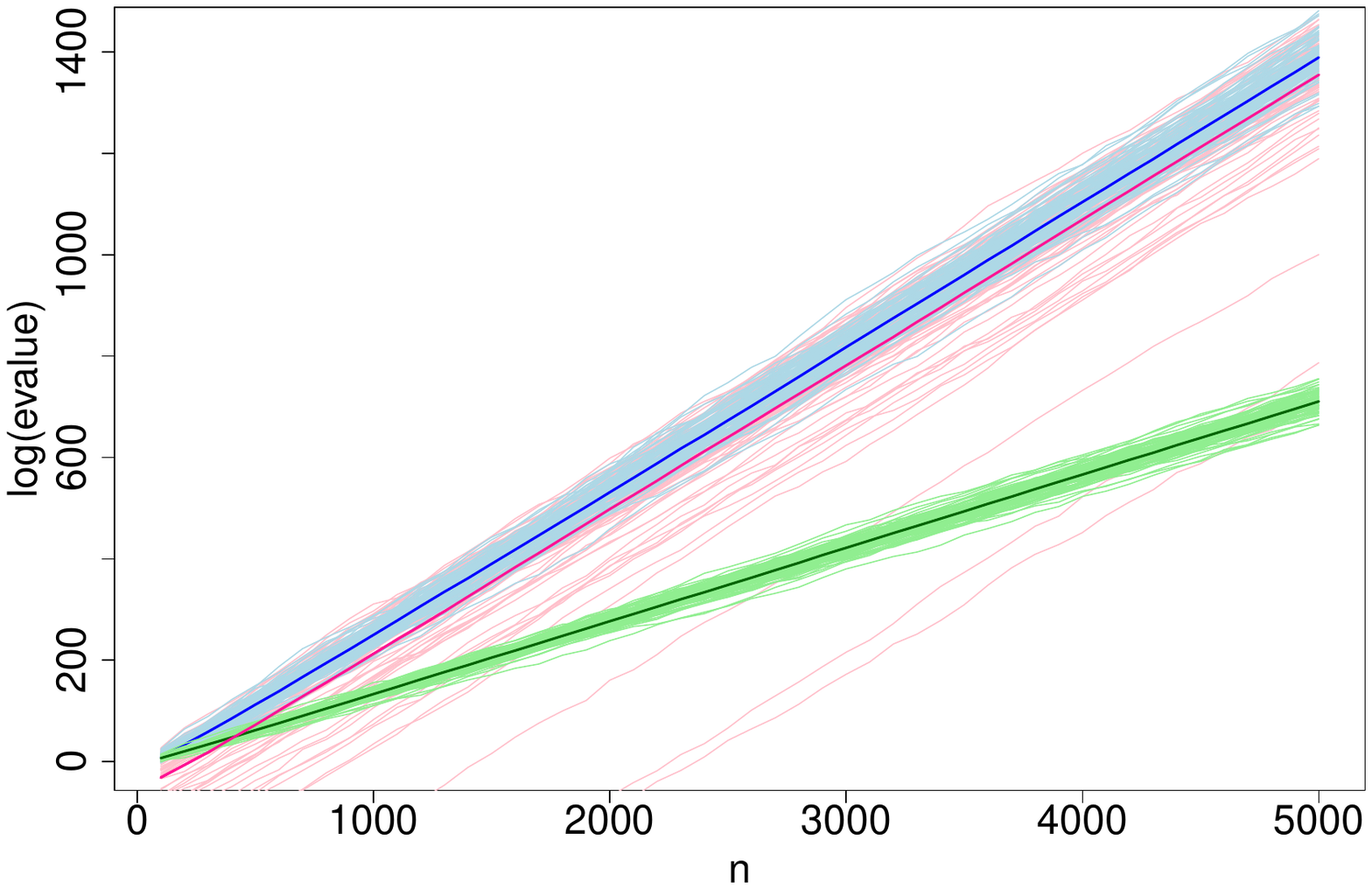}
\label{fig:logconcave2}}
\subfigure[$\mu=14$]
{\includegraphics[width = 0.315\linewidth]{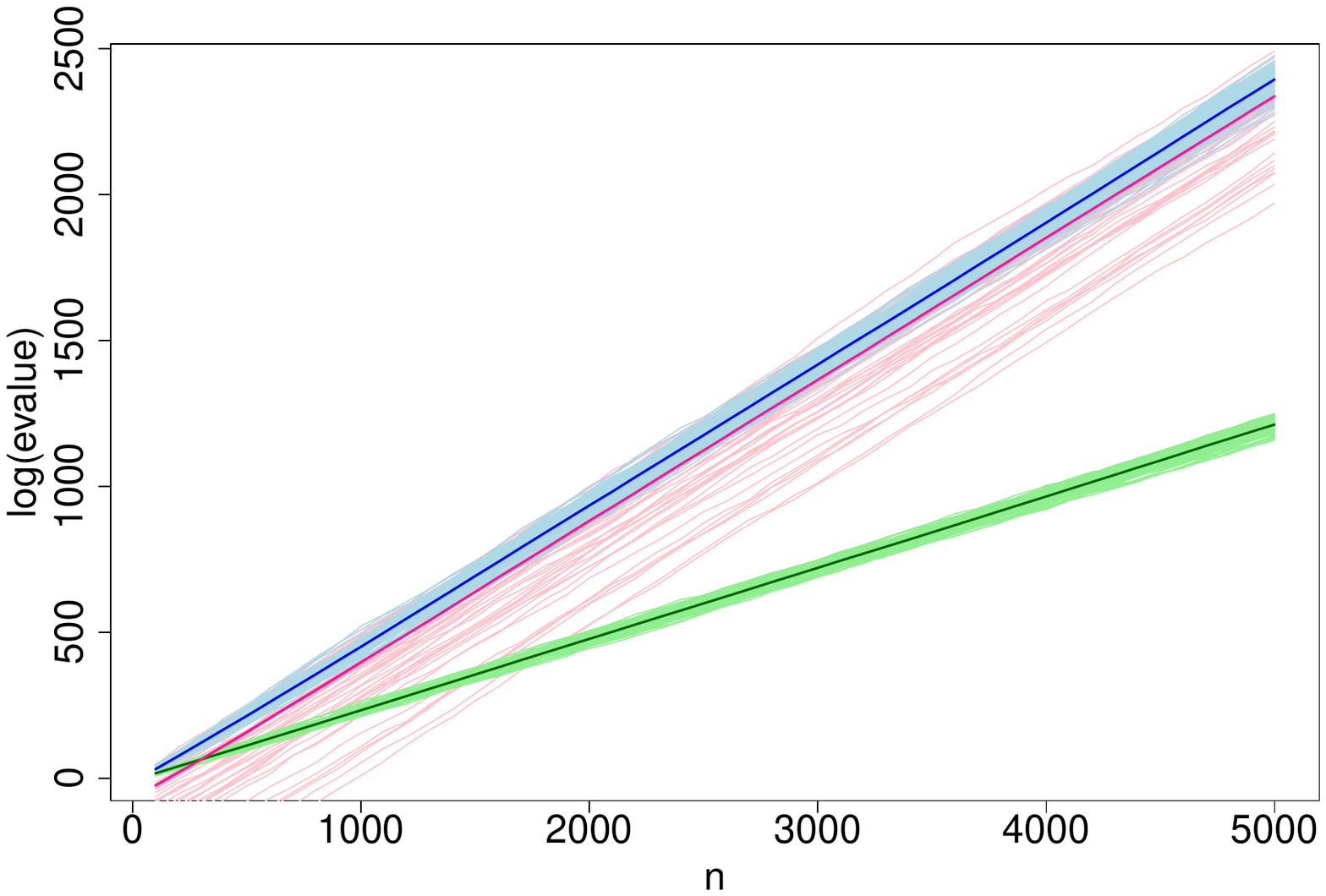}
\label{fig:logconcave3}}
\caption{Plots of $\log E_n^\text{\sc pr}$ (blue), $\log E_n^\text{\sc ulr}$ (pink) and $\log E_n^\text{\sc ui}$ (green) versus $n$, based on the not-log-concave density $p^\star$ as a mixture of normals. Panel~(a) corresponds to $p^{\star}$ ``closest'' to log-concave while (c) is ``farthest.''}
\label{fig:logconcave}
\end{figure}

For the simulation, we take $P^\star$ as an uneven mixture of two normals, with means $0$ and $\mu$ and variances 2. As above, $\mu$ also acts as a measure of the ``degree of non-log-concavity,'' i.e., if $\mu$ is close to 0, then $p^\star$ is only mildly non-log-concave; otherwise, $p^\star$ is more severely non-log-concave.  The three cases we consider in our experiments are $\mu = 6,10,14$. We generate 100 data sets under each, calculate $E_n^\text{\sc pr}$, $E_n^\text{\sc ui}$ and $E_n^\text{\sc ulr}$ at increments $n=100, 200, \ldots, 5000$. Plots of $\log E_n$ versus $n$, along with a pointwise average over replications for each method are shown in Figure~\ref{fig:logconcave}.  The key takeaways are as follows.  First, in terms of statistical efficiency, our $\pr$e-process's growth rate is faster than Dunn et al.'s and no slower than Gangrade et al.'s in each scenario.  Interestingly, the slopes of the $n$ versus $\log E_n^\text{\sc ui}$ lines in Figure~\ref{fig:logconcave} are roughly half that of the $n$ versus $\log E_n^\text{\sc pr}$ lines, which suggests that the loss of efficiency is due to the 50--50 data-splitting Dunn et al.~employ.  Second, in terms of stability, the log-$\pr$e-process has considerably smaller variance than Gangrade et al.'s log-e-process.  Third, in terms of computational efficiency, both $E_n^\text{\sc ui}$ and $E_n^\text{\sc ulr}$ process the entire observed data sequence each time a batch of data arrives because the previous calculation cannot be directly updated.  The $\pr$e-process, on the other hand, has a genuinely recursive update and is much faster. 

The above comparison focused on the scalar-data case, but the same e-processes can be defined and used for testing log-concavity in multivariate settings.  In the supplementary material we present a comparison of $E_n^\text{\sc pr}$ and $E_n^\text{\sc ui}$ in a multivariate case.  The take-away message there is that increasing the dimension naturally impacts the quality of the tests, but it affects the $\pr$e-process less in the sense that $E_n^\text{\sc pr}$ tends to be closer to a suitable oracle e-process than $E_n^\text{\sc ui}$ does.   



\subsection{Testing time homogeneity}
\label{SS:markov}

Let $X_i = (X_{i,1}, \ldots, X_{i,T})$ denote a binary time series of length $T \geq 1$, i.e., $X_{i,t} \in \{0,1\}$ for each $(i,t)$ pair, generally dependent within the $X_i$'s, but independent between $X_i$ and $X_j$, for $j \neq i$.  This situation arises when a binary characteristic, such as employed/unemployed, is measured on subject $i$ at time $t$.  \citet[][Sec.~6]{newtonzhang} consider an application where $X_{i,t}=1$ if dairy cow $i$'s milk sample at time $t$ tests positive for pathogens and $X_{i,t}=0$ otherwise.  The relevant question here is whether the time series are homogeneous or, in other words, is there temporal dependence within the time series or are they independent?  Below we construct a $\pr$e-process and anytime valid test for independence versus a type of Markovian dependence. 

Under the null hypothesis of time homogeneity, each $X_i$ consists of a sequence of Bernoulli trials with its own success probability $\omega_i$.  In this case, the likelihood is binomial, which, for data $X^n = (X_1,\ldots,X_n)$, is maximized at $\hat\omega_i = T^{-1} \sum_{t=1}^T X_{i,t}$, for $i=1,\ldots,n$.  This maximum likelihood goes in the denominator of the $\pr$e-process.  For the alternative, we consider a mixture of Markov models where the mass function $q^\Psi(x)$ is given by \eqref{eq:mixture} with kernel 
\[ p_u(x) = u_{00}^{t_{00}(x)} \, (1-u_{00})^{t_{01}(x)} \, (1-u_{11})^{t_{10}(x)} \, u_{11}^{t_{11}(x)}, \quad x=(x_1,\ldots,x_T), \]
$t_{ab}(x)$ the number of transitions in series $x$ from state $a \in \{0,1\}$ to state $b \in \{0,1\}$, and $u=(u_{00}, u_{11}) \in \UU = [0,1]^2$ is the pair of probabilities corresponding to the ``transitions'' $0 \to 0$ and $1 \to 1$, respectively.  As described in Section~\ref{S:prevalues}, we apply the $\pr$ algorithm to fit this mixture model and obtain likelihood $\hat q^\text{\sc pr}(X^n)$, which is goes in the $\pr$e-process's numerator.  

\begin{figure}[t]
\centering
\subfigure[Null is true]
{\includegraphics[width = 0.4\linewidth]{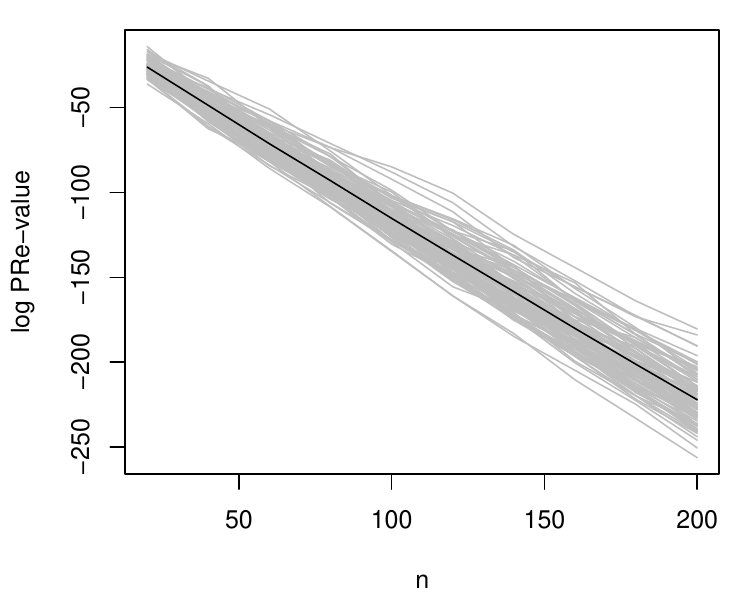}}
\subfigure[Alternative is true]
{\includegraphics[width = 0.4\linewidth]{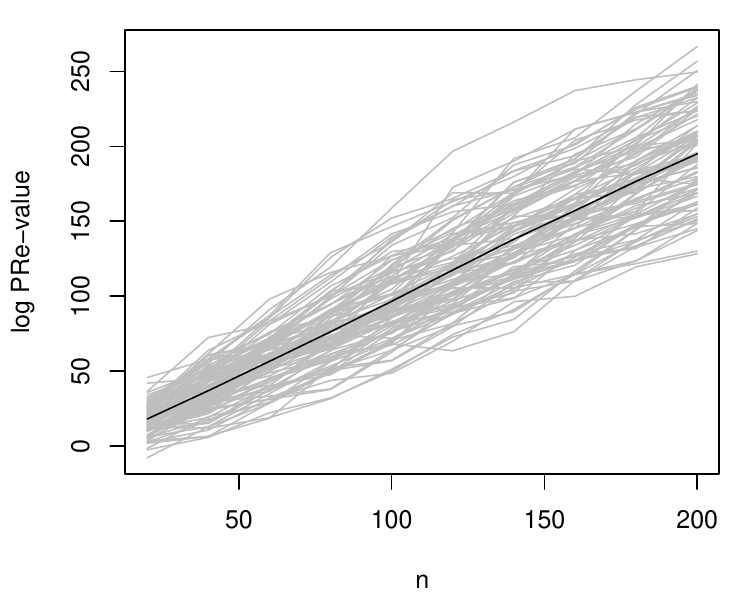}}
\caption{Plots of the $\pr$e-process trajectories, $\log E_n^\text{\sc pr}$ versus $n$, for the time homogeneity tests in Section~\ref{SS:markov}.}
\label{fig:markov}
\end{figure}

For illustration, we consider two simulation settings.  The first is where the null is true, with each $X_i$ being a sequence of Bernoulli trials with success probability $\omega_i \sim \unif(0,1)$.  The second is where the alternative is true, so that $X_i$ is a Markov chain and its transition probability matrix is determined by the pair $(U_{00,i}, U_{11,i})$ iid $\bet(2, 4)$. In both cases, the processes are of length $T=15$, and sample sizes are $n=20, 40, \ldots, 200$.  The paths of $\log E_n^\text{\sc pr}$ versus $n$ for the null and alternative cases are shown in Figure~\ref{fig:markov} and, as desired, we find that $E_n^\text{\sc pr}$ vanishes rapidly under the null and diverges similarly rapidly towards infinity under the alternative.

\section{Conclusion}
\label{S:discuss}

This paper proposes a universal inference-inspired e-process construction based on the $\pr$ algorithm---a fast, recursive procedure for fitting flexible, nonparametric mixture models.  We show that $\pr$e-process is a genuine e-process and, therefore, offers provable, finite-sample, anytime valid inference.  Further, we show that the $\pr$e-process attains the optimal first-order growth rate under the alternative relative to the mixture model fit by $\pr$.  Numerical results demonstrate that the $\pr$e-process's finite-sample empirical performance closely agrees with the optimal behavior predicted by the asymptotic theory.  Moreover, here and in the supplementary material, our proposal with a $\pr$ fit in the e-process numerator is shown to be computationally more efficient, numerically more stable, and more accurately approximates the oracle e-process compared to, say, a kernel density fit in the numerator.  

The $\pr$e-process developments here shed new light on $\pr$ and motivates further computational and theoretical investigations.  In particular, there are open questions concerning $\pr$'s convergence rates (as it pertains to our $\pr$e-process's growth rate here and otherwise) and a need for finite-sample bounds and efficient numerical methods for handling higher-dimensional multivariate data.  On the last point, our $\pr$ticle filtering procedures employed here for multivariate problems is able to accommodate mixtures over moderate-dimensional space $\UU$ but, when the data and hence $\UU$ are high-dimensional, we do not expect the $\pr$ticle filter to be especially accurate, at least not in a reasonable time, so new ideas are needed to tackle those cases.


\section*{Acknowledgments}

Thanks to Aaditya Ramdas for feedback on and corrections to a previous version, and to the other anonymous reviewers for their very helpful suggestions.  This work is partially supported by the U.S.~National Science Foundation, SES--2051225.

\bibliographystyle{apalike}
\bibliography{mybib}

\newpage
\appendix

\begin{center}
{\bf \Large Supplementary Material}
\end{center}

\bigskip

This supplementary material contains some additional technical details that could not fit in the main paper.  In particular, Section~1 offers some new perspectives on the predictive recursion ($\pr$) algorithm in terms of Dawid's prequential statistics; Section~2 offers some further details on the new $\pr$e-process proposed in the main paper; Section~3 gives a detailed statement of the conditions needed to analyze the $\pr$ estimates, along with a proof of the asymptotic growth rate result (Theorem~2) in the main paper; Section~4 present some old and new details on the rate of convergence of $\pr$'s estimator; Section~5 offers some empirical results to help pin down the magnitude of the ``$o(n)$'' error term in Theorem~2's asymptotic approximation; Section~6 provides a comparison of two e-processes in multivariate settings; and Section~7 proposes a variation on the $\pr$e-process presented in the main paper.

\section{New perspectives on PR}

Both of us wrote our PhD theses on predictive recursion ($\pr$), viewed as an estimation procedure.  That is, the focus of those previous efforts was on large-sample consistency of $\pr$'s estimated mixing and mixture distributions and on applications in high-dimensional inference.  The present paper takes a different perspective, focusing on $\pr$ in the context of {\em prediction}, which, ironically, has not been thoroughly investigated in the literature.  This is both interesting and relevant to the main paper, so we collect some of these developing thoughts---potential new lines of research---here in this supplementary material.  We specifically would like to thank one of the anonymous reviewers for pointing out some of these connections.  

First, there are a number of methods available in the machine learning literature for online or sequential prediction, i.e., algorithms that produce a sequence of probability distributions $(Q_t)$, $t = 1, 2, \ldots$, where $Q_t$ is interpreted as the predictive distribution of $X_{t+1}$ depending on $X^t$.  Like with $\pr$, it is not uncommon to construct these predictive distributions using mixtures.  For example, in \citet{devaine.etal.2013}, the authors consider applications where there is a fixed, finite number of of experts making forecasts about $X_{t+1}$ based on previous data $X^t$, and the proposed algorithm aggregates the experts' forecasts using mixtures; see also \citet{littlestone.warmuth.1994} and \citet{bousquet.warmuth.2002}.  \citet{devaine.etal.2013} go on to prove non-asymptotic bounds on the regret---prediction performance compared to an oracle---for certain convex loss functions.  While the particular algorithms in the above references are different from $\pr$, this connection is relevant for at least two reasons: first, any sequential aggregation algorithm ought to have its own corresponding e-process---an {\sc aggre}e-process, say?---that warrants investigation, just like $\pr$ and our proposed $\pr$e-process; second, perhaps the theoretical analysis of these alternative algorithms can shed light on, say, how to establish similar regret bounds for $\pr$.  

Next, and particularly relevant to the details in the present paper, is a connection to Dawid's {\em prequential} approach to and perspective on probability and statistics \citep[e.g.,][]{dawid1984, dawid1991}.  Even the names are similar: the ``prequential'' is a combination of the words ``predictive'' and ``sequential,'' which clearly has aspects in common with ``predictive recursion.''  A key difference, however, is that what Dawid has put forward is a general theory or perspective, while $\pr$ provides a particular instantiation of it, as we now explain.  The sequence $(Q_t)$ discussed in the previous paragraph is what Dawid calls a {\em prequential forecasting system}, which could take all sorts of different forms, e.g., from a fully specified joint distribution for the observables to a more ad hoc expert aggregation strategy.  $\pr$ falls somewhere in between these two extremes: $\pr$ has at its core a flexible, nonparametric mixture model for the observables, but it learns from observations and updates in a non-Bayesian/non-probabilistic fashion.  Dawid argues, as formulated in his {\em prequential principle}, that the quality of a forecasting system should be determined solely based on how well it predicts the actual observations; in other words, this assessment should not take into consideration hypothetical outcomes that were not observed.  Our proposed $\pr$e-process test in the main paper is exactly in this spirit.  Indeed, just as in \citet[][Sec.~7]{dawid1984}, our $\pr$e-process in Equation~(6) in the main text, is the ratio of two forecasting systems, each applied to the observed data sequence: the numerator represents the $\pr$-based forecasts and the denominator a summary of the forecasts under the null model $\model_0$.  Then the proposed test rejects the null model if and only if the ratio exceeds some specified threshold (which does not depend on, say, ``null distribution'' characteristics, etc).  In other words, the null model is rejected if and only if the $\pr$-based forecasting system is significantly ``better'' than that which assumes the null model.  That our $\pr$e-process test is consistent with the prequential principle itself is not enough to justify its use.  It is precisely for this reason that Theorems~1--2 in the main paper are essential: on the one hand, the $\pr$e-process $(E_n^\text{\sc pr})$ stays small forever when $\model_0$ is true and, on the other hand, (under certain conditions) it diverges quickly to infinity when $\model_0$ is not true.  

\citet{dawid1985} discusses the use of forecasting systems to define an objective, empirical notion of probability assigned to observed data sequences, and comments on the implications to the foundations of statistics, probability, and science more broadly; see, also, \citet{vovk1993}.  Not just any forecasting system achieves this objective, and Dawid highlights a key {\em calibration} condition.  Using our present notation, fix a sequence $(A_n)$ of subsets of $\XX$ and the observations $a_n = 1(X_n \in A_n)$, where $1(\cdot)$ is the indicator.  Sparing some technical details---concerning computable selection rules, essential to properly defining ``randomness'' in the spirit of von Mises---a forecasting system $(Q_n)$ is {\em calibrated} with respect to the observed sequence $(a_n)$ if 
\[ \frac1n \sum_{t=1}^n \{ Q_t(A_t) - a_t \} \to 0, \quad n \to \infty. \]
This has important implications, namely, that any two forecasting systems that are calibrated to $(a_n)$ must be asymptotically equivalent \citep[e.g.,][]{blackwell.dubins.1962}, and it is precisely that common ``asymptotic probability'' that Dawid puts forward as the empirical probability of $(a_n)$.  For the $\pr$ forecasting system $(\widehat Q_n^\text{\sc pr})$, if $X_1,X_2,\ldots$ are iid $P^\star$, then 
\begin{align*}
\frac1n \sum_{t=1}^n \{ \widehat Q_t^\text{\sc pr}(A_t) - a_t \} & = \frac1n \sum_{t=1}^n \{ \widehat Q_t^\text{\sc pr}(A_t) - P^\star(A_t) + P^\star(A_t) - a_t \} \\
& = \frac1n \sum_{t=1}^n \{\widehat Q_t^\text{\sc pr}(A_t) - P^\star(A_t)\} + \frac1n \sum_{t=1}^n \{ P^\star(A_t) - a_t \}. 
\end{align*}
It follows from the standard law of large numbers for bounded, mean zero random variables that the second term in the last line above converges to 0 with $P^\star$-probability 1.  If $P^\star$ is contained in the set $\convex$ defined in Section~3.3 of the main paper, and if the conditions in Section~3 below are satisfied, then it follows from Corollary~4.6 in \citet{mt-rate} that $\widehat Q_n^\text{\sc pr} \to P^\star$ with $P^\star$-probability 1 in total variation. This implies that the first term in the above display is vanishing as $n \to \infty$ and, consequently, the $\pr$ forecasting system is calibrated with respect to the sequence $a_n = 1(X_n \in A_n)$ determined by $X^\infty$, for any such $X^\infty$ in a set of $P^\star$-probability 1.  This, of course, is an especially simple case; but $\pr$ itself is quite flexible, e.g., it does not make use of the iid assumption, so we suspect that calibration can be established more generally. 

Finally, under this umbrella of forecasting systems, the idea put forward in \citet{mt-prml} can be given a new, potentially useful twist.  They proposed a semiparametric mixture model wherein the kernel $\{p_u: u \in \UU\}$ used to the define the $\pr$ algorithm depends on some parameter $\theta \in \TT$; that is, their kernel is written as $p_{u,\theta}$, with $x \mapsto p_{u,\theta}(x)$ a probability density function for each $(u,\theta)$.  At any fixed value of $\theta$, we can produce $\pr$'s joint marginal density for $X^n$, denoted here as $\hat q_\theta^\text{\sc pr}(X^n)$.  Then \citet{mt-prml} defined the ``likelihood'' function $\theta \mapsto \hat q_\theta^\text{\sc pr}(X^n)$ and compared it to the marginal likelihood function under a Bayesian nonparametric mixture model with a Dirichlet process prior.  They then proposed an estimate of $\theta$ based on maximizing that $\pr$-marginal likelihood: $\hat\theta_n^\text{\sc pr} = \arg\max_\theta \hat q_\theta^\text{\sc pr}(X^n)$.  Unfortunately, despite ample empirical evidence suggesting it, sufficiently general results on large-sample consistency of $\hat\theta_n^\text{\sc pr}$ have yet to be established.  But one can interpret Martin \& Tokdar's suggestion as a particular instantiation of the ``prequential model'' in \citet[][Sec.~7.3]{dawid1991} which, for each $\theta$, evaluates forecasts for the next observation in light of what has already been observed.  If, as conjectured in the previous paragraph, this $\pr$-based prequential model can be shown to have high-level justification in terms of calibration, etc., then it would be meaningful to assess the properties of $\hat\theta_n^\text{\sc pr}$ relative to the $\pr$-based model.  This creates a new and different opportunity to demonstrate the utility of this estimator.  Indeed, \citet{dawid1984, dawid1991} presents a notion of (simple) prequential consistency, in which the predictions of $X_{n+1}$ based on $\widehat Q_{\hat\theta_n^\text{\sc pr}}^\text{\sc pr}$ and $\widehat Q_{\theta^\star}^\text{\sc pr}$, given $X^n=x^n$, are compared as $n \to \infty$, where $\theta^\star$ denotes the ``true'' value in the $\pr$-based prequential model.  If these predictions are asymptotically equivalent, then Dawid would call $\hat\theta_n^\text{\sc pr}$ (simply and prequentially) {\em consistent}.  Aside from the attractiveness of the prequential perspective, we believe that the aforemention demonstration is within reach.  This is also promising too, as Dawid argues, this notion of consistency often holds more broadly than the classical notions of consistency.

\section{Further PRe-process details}

The paper's main text describes and focuses on the use of our proposed $\pr$e-process $E_n^\text{\sc pr}$ for anytime valid hypothesis testing.  But there is more that can be done with the $\pr$e-process and here we briefly describe three such developments.  

First, given a null model, $\model_0$, one can define a corresponding ``anytime p-value'' 
\[ \pi^\text{\sc pr}(X^n; \model_0) = 1 \wedge E^\text{\sc pr}(X^n; \model_0)^{-1}. \]
Just like in Corollary~1 in the main paper, it follows that
\[ \sup_{P \in \model_0} P\{ \pi^\text{\sc pr}(X^N; \model_0) \leq \alpha \} \leq \alpha, \quad \text{all $\alpha \in [0,1]$ and all stopping times $N$}. \]
So, as expected, the anytime valid test proposed in the main paper is equivalent to a test that rejects when the above p-value is no more than $\alpha$.   

Second, by considering singleton null hypotheses $\model_0 = \{P_0\}$ and varying $P_0$, the above p-value defines a function $P_0 \mapsto \pi^\text{\sc pr}(X^n; \{P_0\})$ which, as explained in \citet{mycomment.ramdas2023}, determines a data-dependent imprecise probability distribution supported on $\model$, in particular, a necessity--possibility measure pair \citep[e.g.,][]{dubois.prade.book, shafer1976}.  This, in turn, determines an anytime valid {\em inferential model} \citep{imbasics, imbook, martin.partial2} that provides provably safe and reliable imprecise-probabilistic uncertainty quantification about the unknown $P$.  Akin to a Bayesian posterior distribution, this imprecise-probabilistic uncertainty quantification can then be used, e.g., for formal decision-making \citep{imdec}.  Indeed, if $(a,P) \mapsto \ell_a(P)$ quantifies the loss of taking action $a$ when the state of the world is $P$, then one can find a lower and upper expected loss using a specialization of Choquet's general theory of integration with respect to capacities \citep{choquet1953} to the case of necessity and possibility measures \citep[e.g.,][Sec.~7.8]{lower.previsions.book}.  From here, one can proceed to, say, minimize the upper expected loss with respect to the action $a$ to obtain an inferential model version of a Bayes rule, and establish bounds on the difference between the inferential model's upper expected loss and the oracle loss that knows the true $P$.  Since this is not specific to the $\pr$e-process developments in this paper, we will not go into any further details here.  But the interested reader might like to see \citet{denoeux.decision.2019} for a review of the literature on this brand of generalized Bayes decision theory, and also R.~Martin's extended abstract in the forthcoming Oberwolfach Mathematical Research Institute report (\url{https://www.mfo.de/occasion/2419b}). 

Finally, a more familiar consequence of the anytime validity of the $\pr$e-process-based test in the main paper is inverting the test to construct an anytime valid confidence set.  Let $\phi: \model \to \Phi$ be a map that extracts some relevant feature from $P \in \model$.  For example, $\phi(P) = \int_\XX x \, P(dx)$ is the mean of $P$, $\phi(P) = \inf\{x: P((-\infty,x]) \geq \tau\}$ is the $\tau^\text{th}$ quantile of $P$, and $\phi(P)=P$ is the distribution $P$ itself.  For a given $\phi$, define the following data-dependent subset of $\Phi$:
\[ C_\alpha(X^n) = \bigl\{ \phi(P): \pi^\text{\sc pr}(X^n; \{P\}) > \alpha \bigr\}, \quad \alpha \in [0,1]. \]
The set $C_\alpha(X^n)$ is just the collection of all $\phi(P)$ corresponding to ``null'' $P$s that the $\pr$e-process-based test would not reject for data $X^n$ at level $\alpha$.  Then the fact that $E_n^\text{\sc pr}$ is an e-process implies that $C_\alpha(\cdot)$ is an anytime valid $100(1-\alpha)$\% confidence set for $\phi(P)$ in the sense that 
\[ \sup_{P \in \model} P\{ C_\alpha(X^N) \not\ni \phi(P) \} \leq \alpha, \quad \text{any $\alpha \in (0,1)$ and any stopping time $N$}. \]
Of course, identifying the set $C_\alpha(X^n)$ for a given $\phi$ and a given data set $X^n$ could be a challenge, but approximations could surely be found if there was sufficient desire to do so. 


\section{The PRe-process's asymptotic growth rate}
\label{A:proof}

\subsection{Setup of Theorem~2}
\label{AA:setup}

There are a few inputs that play key roles in the asymptotic properties of the $\pr$ algorithm.  These include some user-specified inputs, namely, the family of kernel densities $\{p_u: u \in \UU\}$ that determine the set of mixtures $\convex$, and $\pr$'s weights $(w_i)$ and initial guess $\Psi_0$; two more relevant quantities that are beyond the user's control are the true distribution $P^\star$, with density $p^\star$, and the corresponding Kullback--Leibler projection of $P^\star$ onto the set $\convex$ of mixtures.  This latter projection is defined as the distribution $Q^\star$ with density $q^\star$ that satisfies 
\begin{equation}
\label{a:eq:inf}
K(P^\star, Q^\star) = K(P^\star, \convex) := \inf_{Q \in \convex} K(P^\star, Q). 
\end{equation}
The existence of $Q^\star$ is ensured by various sets of conditions; see, e.g., \citet[][Ch.~8]{liesevadja}.  In particular, existence of $Q^\star$ is implied by Condition~\ref{cond:support} below, as shown in Lemma~3.1 of \citet{mt-rate}.  The following conditions, on which Theorem~2 is based, concern the basic properties of these individual inputs and their interplay. 

\begin{condition}
\label{cond:support}
$\UU$ is compact and $u \mapsto p_u(x)$ is continuous for almost all $x$. 
\end{condition}

\begin{condition} 
\label{cond:weights}
The $\pr$ weight sequence $(w_i)$ satisfies 
\begin{equation}
\label{eq:weights}
\sum_{i=1}^\infty w_i = \infty \quad \text{and} \quad \sum_{i=1}^\infty w_i^2 < \infty. 
\end{equation}
\end{condition}

\begin{condition}
The kernel density $p_u(x)$ satisfies
    \label{cond:integrable}
    \begin{equation}
    \sup_{u_1, u_2 \in \UU} \int \Bigl \{\frac{p_{u_1}(x)}{p_{u_2}(x)} \Bigr \}^{2} \, p^\star(x) \, dx < \infty
    \end{equation}
\end{condition}



\begin{condition}
\label{cond:condition4}
The Kullback--Leibler projection $Q^\star$ in \eqref{a:eq:inf}, with density $q^\star$, satisfies
\begin{equation}
\label{eq:condition4}
\int \Bigl\{ \log \frac{p^\star(x)}{q^\star(x)} \Bigr\}^2 \, p^\star(x) \, dx < \infty. 
\end{equation}    
\end{condition}

Here we offer some explanation and intuition.  First, compactness of the mixing distribution support $\UU$ in Condition~\ref{cond:support} is difficult to relax, but the fact that $\UU$ can be taken arbitrarily large means that this imposes effectively no practical constraints on the user.  Continuity of the kernel can be relaxed, but at the expense of a much more complicated condition; the reader interested in this can consult Equation~(7) in \citet{dixit.martin.revisiting} and the relevant discussion.  Condition~\ref{cond:weights} says simply that the weights must be vanishing to ensure convergence but not too quickly since the algorithm needs an opportunity to learn; the requirement in \eqref{eq:weights} is just right for this. Condition~\ref{cond:integrable} is non-trivial but holds for Gaussian and other exponential family distributions thanks to the compactness of $\UU$ in Condition~\ref{cond:support}.  
Finally, Condition~\ref{cond:condition4} concerns the quality of the mixture model itself.  One cannot hope to achieve quality estimation/inference in any sense if the ``best'' member of the mixture model differs considerably from the true density $p^\star$.  Equation \eqref{eq:condition4} is just a particular way to say that $p^\star$ and $q^\star$ do not differ by too much.  

Once the user makes his/her specification of the mixture model, Condition~\ref{cond:integrable} determines a set of true densities $p^\star$ for which the $\pr$ algorithm will provide consistent estimation.  Indeed, Theorem~1 in \citet{mt-prml} states that, under Conditions~\ref{cond:support}--\ref{cond:integrable}, the $\pr$ estimator $\hat q_{X^n}$ satisfies $K(p^{\star}, \hat q_{X^n}) \to \inf_{q \in \convex} K(p^{\star}, q)$ with $P^\star$-probability~1 as $n \to \infty$.  The user, of course, can vary the mixture model specification to tailor $\pr$ toward what they expect $p^\star$ to look like.  But we need more than consistency for our purposes here, and Conditions~\ref{cond:condition4} further restricts the set of true densities to those for which the $\pr$ algorithm can give us the ``more'' that we need.



\subsection{Proof of Theorem~2}

As explained in the main text, Theorem~2 is a consequence of the following property of $\pr$:
\begin{equation}
\label{eq:pr.limit}
n^{-1} \log \hat q^\text{\sc pr}(X^n) = n^{-1} \log p^\star(X^n) - K(P^\star, \convex) + o(1), \quad n \to \infty, 
\end{equation}
The goal here is simply to show that \eqref{eq:pr.limit} holds under Conditions~\ref{cond:support}--\ref{cond:condition4} as stated above.  The argument closely follows that in \citet{mt-prml}, but we provide the details here for completeness since their context and notation is different from ours. 

The strategy of the proof is as follows.  First, simplify the notation by writing $\hat q_{i-1}(X_i) = \hat q_{X^{i-1}}(X_i)$ for each $i$.  Next, define the sequence of random variables 
\[K_n = \frac{1}{n}\sum_{i=1}^{n} \log {\frac{p^{\star} (X_i)}{\hat q_{i-1}(X_i)}}, \]
which might be interpreted as a sort of empirical Kullback--Leibler divergence.  If $K^{\star} = K(P^\star, \convex)$, then \eqref{eq:pr.limit} is equivalent to $K_n \to K^\star$ with $P^{\star}$-probability~1 as $n \to \infty$.  It is this latter claim that we will prove below, using a martingale strong law in \citet{teicher98}.  

Towards this, define a sequence of random variables $Z_i$ as,
\[Z_i = \log {\frac{p^{\star} (X_i)}{\hat q_{i-1}(X_i)}} - K(p^\star, \hat q_{i-1}) , \quad i\geq 1. \]
Recall that $\mathscr{A}_{i-1} = \sigma(X^{i-1})$, so $E_{P^\star}(Z_i \mid \mathscr{A}_{i-1}) = 0$ and, therefore, $\{(Z_i, \mathscr{A}_{i}): i \geq 1\}$ is a zero mean martingale sequence under $P^\star$. For $q^{\star}$ as in \eqref{a:eq:inf}, we have 
\begin{align*}
    E_{P^\star}(Z_i^2 \mid \mathscr{A}_{i-1}) &\leq \int \left\{\log \frac{p^{\star} (x)}{\hat q_{i-1} (x)}\right\}^{2} p^{\star}(x) \, dx\\
    &= \int \left\{\log \frac{q^{\star} (x)}{\hat q_{i-1} (x)} + \log \frac{p^{\star} (x)}{q^{\star} (x)} \right\}^{2} p^{\star}(x) \, dx\\
    &\leq 2 \int \left\{\log \frac{q^{\star} (x)}{\hat q_{i-1} (x)}\right\}^{2} p^{\star} (x) \, dx + 2 \int \left\{\log \frac{p^{\star} (x)}{q^{\star} (x)} \right\}^{2} p^{\star}(x) \, dx\\
    &= 2 A_i + 2 B.
\end{align*}
Of course, $B$ is a finite constant according to Condition~\ref{cond:condition4}. To bound $A_i$ let us first define $\XX_0 = \{x : q^{\star} (x) < \hat q_{i-1} (x)\}$. Using basic properties of the logarithm, we get
\begin{align*}
    A_i &= \int \left\{\log \frac{q^{\star} (x)}{\hat q_{i-1} (x)}\right\}^{2} p^{\star} (x) \, dx\\
    &= \int_{\XX_0} \left\{\log \frac{\hat q_{i-1} (x)}{q^{\star} (x)}\right\}^{2} p^{\star} (x) \, dx + \int_{\XX_0^c} \left\{\log \frac{q^{\star} (x)}{\hat q_{i-1} (x)}\right\}^{2} p^{\star} (x) \, dx\\
    &\leq \int_{\XX_0} \left\{\frac{\hat q_{i-1} (x)}{q^{\star} (x)} - 1\right\}^{2} p^{\star} (x) \, dx + \int_{\XX_0^c}\left\{\frac{q^{\star} (x)}{\hat q_{i-1} (x)} - 1 \right\}^{2} p^{\star} (x) \, dx\\
    &\leq 2 + \int\left[ \left\{\frac{\hat q_{i-1} (x)}{q^{\star} (x)}\right\}^{2} + \left\{\frac{q^{\star} (x)}{\hat q_{i-1} (x)}\right\}^{2} \right] p^{\star} (x) \, dx.
\end{align*}
Since both $\hat q_{i-1}$ and $q^\star$ in the two numerators in the above display are mixtures of the kernel $p_u$, we can say that
\[A_i \leq 2 + 2 \sup_{u_1, u_2 \in \UU} \int \Bigl \{\frac{p_{u_1}(x)}{p_{u_2}(x)} \Bigr \}^{2} \, p^\star(x) \, dx\]
which is bounded by Condition~\ref{cond:integrable}. Therefore, $E_{P^\star}(Z_i^2 \mid \mathscr{A}_{i-1})$ is bounded too.
Switching from index ``$i$'' to the more natural ``$n$,'' we have that 
\[\frac{E_{P^\star}(Z_n^2 \mid \mathscr{A}_{n-1})}{n^2 (\log \log n)^{-1}} \lesssim n^{-2} (\log \log n) \to 0. \]
Therefore, by Markov's inequality, with $P^\star$-probability~1 we have,
\[ \sum_{n=1}^{\infty} P^\star\Bigl( |Z_n| > \frac{n}{\log \log n} \bigmid \mathscr{A}_{n-1} \Bigr) \lesssim \sum_{n=1}^{\infty} \frac{(\log \log n)^{2}}{n^2} < \infty. \]
From this and Corollary~2 of \citet{teicher98}---with his ``$\beta=1$''---we get $n^{-1}\sum_{i=1}^{n} Z_i \to 0$ with $P^\star$-probability~1. Therefore, also with $P^\star$-probability~1,
\[ \Bigl| K_n - \frac{1}{n} \sum_{i=1}^{n} K(p^{\star}, \hat q_{i-1}) \Bigr| = \Bigl| (K_n - K^{\star}) - \frac{1}{n} \sum_{i=1}^{n} \{K(p^{\star}, \hat q_{i-1}) - K^{\star}\} \Bigr| \to 0. \]
From Theorem~1 in \citet{mt-prml} and Cesaro's theorem, we have $K_n - K^{\star} \to 0$ with $P^\star$-probability~1.

\subsection{Condition (9) concerning the null model}

Theorem~2 in the main paper---see Equation~(9)---relies on a particular feature $\kappa^\star(\model_0)$ of the null model $\model_0$, , namely, 
\begin{equation}
\label{eq:mle.limit}
\liminf_{n \to \infty} n^{-1} \log\{ p^\star(X^n) / \hat p_0(X^n) \} \geq \kappa^\star(\model_0), \quad \text{with $P^\star$-probability 1}, 
\end{equation}
where $\hat p_0(X^n) = \sup_{p_0 \in \model_0} p_0(X^n)$ is the maximum likelihood in $\model_0$.  As stated in the main paper, it is easy to check that $\kappa^\star(\model_0)$ is bounded above by $K(P^\star, \model_0)$.  Indeed, by definition of maximum likelihood,  
\[ n^{-1} \log\{ p^\star(X^n) / \hat p_0(X^n) \} \leq n^{-1} \log\{ p^\star(X^n) / p_0(X^n) \}, \quad \text{for all $P_0 \in \model_0$}, \]
and the upper bound converges with $P^\star$-probability 1 to $K(P^\star, P_0)$.  This can be made arbitrarily close to $K(P^\star, \model_0)$ through careful choice of $P_0$ and, therefore, $\kappa^\star(\model_0) \leq K(P^\star,\model_0)$.  

Our plan here is to focus on the case where $\kappa^\star(\model_0) = K(P^\star,\model_0)$, but we first want to emphasize that strict inequality is possible.  This corresponds to a particularly unusual case where the maximum likelihood estimator fits the data better, even asymptotically, than the true distribution $P^\star$ (or the model's ``best representative'').  As an illustration, consider the famous Neyman--Scott problem \citep{neyman.scott.1948}, where data $X^n$ is of the form $\{(X_{11}, \ldots, X_{1J}), \ldots, (X_{n1},\ldots,X_{nJ}\}$, fully independent throughout, with $X_{i1},\ldots,X_{in}$ iid $\nm(\mu_i, \sigma^2)$, for $i=1,\ldots,n$.  The point is that there are $n$-many mean parameters $(\mu_1,\ldots,\mu_n)$ but only one variance parameter $\sigma^2$.  The maximum likelihood estimators are 
\[ \hat\mu_i = \frac{1}{J} \sum_{j=1}^J X_{ij}, \quad i=1,\ldots,n \quad \text{and} \quad \hat\sigma^2 = \frac{1}{nJ} \sum_{i=1}^n \sum_{j=1}^J (X_{ij} - \hat\mu_i)^2. \]
What makes this example ``famous'' is that it is one of the first to demonstrate inconsistency of the maximum likelihood estimator.  In particular, if $\sigma^2$ denotes the true variance parameter under $P^\star$, then $\hat\sigma^2 \to \frac{J-1}{J} \sigma^2 < \sigma^2$ with $P^\star$-probability 1 as $n \to \infty$ with $J$ fixed.  In this case, using the general notation from above, it can be shown that 
\[ n^{-1} \log\{ p^\star(X^n) / \hat p_0(X^n) \} \to \log( \tfrac{J-1}{J} ) < 0, \]
while the minimum Kullback--Leibler number for this correctly-specified model is 0.  This happens because maximum likelihood systematically underestimates the variance, and a model with less spread around the observations will tend to fit better in the sense of having non-trivially higher likelihood.  Our hunch is that strict inequality $\kappa^\star(\model_0) < K(P^\star,\model_0)$ holds only when the $\model_0$-maximum likelihood estimator is inconsistent, but we will not pursue this matter here.  

Returning to the general question, we are interested in when 
\[ n^{-1} \log\{ p^\star(X^n) / \hat p_0(X^n) \} \to K(P^\star, \model_0), \quad \text{with $P^\star$-probability 1}. \]
Of course, if the model is correct in the sense that $P^\star \in \model_0$, then the right-hand side is 0 and we would expect that the left-hand would be a limit that also equals 0.  Corollary~3.2 in \citet{doss.wellner.2016} gives a result along these lines for the case of log-concave densities, which was one of the applications in the main paper.  The more interesting case for us is when the model is wrong, i.e., $K(P^\star, \model_0) > 0$. As is common in the literature on model misspecification \citep[e.g.,][]{patilea2001, kleijn}, we assume that there exists $P_0^\dagger \in \model_0$ such that $K(P^\star, P_0^\dagger) = K(P^\star, \model_0)$.   Uniqueness if $P_0^\dagger$ is not necessary for us here, but it is unique if $\model_0$ is convex, e.g., in the monotone density application in the main paper.  Then the scaled log-likelihood ratio can be written as 
\[ n^{-1} \log\{ p^\star(X^n) / \hat p_0(X^n) \} = n^{-1} \log\{ p^\star(X^n) / p_0^\dagger(X^n) \} + n^{-1} \log\{ p_0^\dagger(X^n) / \hat p_0(X^n) \}. \]
The law of large numbers implies that the first term on the right-hand side converges to $K(P^\star, \model_0)$ with $P^\star$-probability 1.  So, only the second term on the right-hand side requires further analysis and, in particular, we would like to know when this would be vanishing as $n \to \infty$.  Since $\hat p_0(X^n) \geq p_0^\dagger(X^n)$, it is enough to show that $n^{-1} \log\{ \hat p_0(X^n) / p_0^\dagger(X^n)\} \to 0$.  

A general strategy is through the use of a suitable uniform law of large numbers.  Consider the following decomposition, with de Finetti's notation $P^\star f$ for expectation $\int f \, dP^\star$, 
\begin{align*}
n^{-1} \log\{ \hat p_0(X^n) / p_0^\dagger(X^n) \} & = \sup_{P_0 \in \model_0} \frac1n \sum_{i=1}^n \{ \log p_0(X_i) - \log p_0^\dagger(X_i)\} \\
& = \sup_{P_0 \in \model_0} \frac1n \bigg[ \sum_{i=1}^n \{ \log p_0(X_i) - P^\star \log p_0\} \\
& \qquad \qquad \quad - \sum_{i=1}^n \{ \log p_0^\dagger(X_i) - P^\star \log p_0^\dagger\} \\
& \qquad \qquad \quad - n\{ K(P^\star, P_0) - K(P^\star, P_0^\dagger) \} \bigg] \\
& \leq \sup_{P_0 \in \model_0} \bigg| \frac1n \sum_{i=1}^n \{ \log p_0(X_i) - P^\star \log p_0\} \bigg| + o(1), 
\end{align*}
where ``$o(1)$'' corresponds to $n^{-1} \sum_{i=1}^n \{ \log p_0^\dagger(X_i) - P^\star \log p_0^\dagger\}$, which is vanishing with $P^\star$-probability 1 by the law of large numbers.  Of course, if $\log \model_0$ is a Glivenko--Cantelli class, then of course the upper bound in the above display is vanishing and the desired result holds.  If, for example, the densities in $\model_0$ are not bounded away from 0, then this naive Glivenko--Cantelli property might fail.  In that case, one can restrict the class $\model_0$ to some better-behaved subset for which the Glivenko--Cantelli property can be established, then show that the maximum likelihood estimator would be in that restricted $\model_0$ with probability converging to 1.  This is relatively standard from here, so we refer the interested reader to the classical texts on empirical process theory and statistical inference/learning \citep[e.g.,][]{vaartwellner1996, vaart1998, vandegeer2000, kosorok.book}.

\section{On PR's rate of convergence}
\label{A:rate}

The rate at which $\pr$'s density estimator converges to the true density $p^\star$ is of general interest and is relevant to questions about the asymptotic growth rate of the $\pr$e-process under the alternative.  Here we review and slightly extend the theoretical analysis in \citet{mt-rate}.  

\citet{mt-rate} showed that, under the conditions of Theorem~2 in the main paper (see Section~\ref{A:proof} above), if $P^\star$ is in the interior of the set $\convex$ of mixtures and if $\pr$'s weight sequence $(w_i)$ has the form $w_i = (i+1)^{-\gamma}$, for $\gamma \in (2/3,1]$, then $\pr$'s density estimator $\hat q_{X^n}$ satisfied $n^{1-\gamma} K(p^\star, \hat q_{X^n}) \to 0$ in $P^\star$-probability.  It is for precisely this reason that we take $\gamma=0.67$, just slightly more than $2/3$, in our numerical examples---it corresponds to roughly the best rate according to the available theory.  Martin \& Tokdar argue that their bound on the rate is conservative or worst-case in the sense that it corresponds to a $P^\star$ arbitrarily close to the boundary of $\convex$; this boundary case is ``worst'' because $\pr$ struggles most to estimate a mixing distribution that does not have a density with respect to the posited dominating measure.  

As an aside, a shortcoming of the theoretical analyses of $\pr$ published to date is that they apparently are not able to accommodate structural assumptions about $P^\star$, so the results all are of a ``worst-case'' form.  This makes it difficult to assess the performance of $\pr$ in natural cases where $P^\star$ has a sufficiently smooth density, etc.  

It turns out that there is another previously-unknown sense in which $\pr$'s convergence rate is conservative.  The following proposition makes this precise.

\begin{proposition}
Under the conditions stated in the second paragraph of this section, 
\[ n^{1-\gamma} \max_{s \geq n} K(p^\star, \hat q_{X^s}) \to 0 \quad \text{in $P^\star$-probability, $n \to \infty$}. \]
\end{proposition}

\begin{proof}
Martin \& Tokdar's proof leans on a demonstration that $K(p^\star, \hat q_{X^n})$ and $n^{1-\gamma} K(p^\star, \hat q_{X^n})$ are what \citet{robbins-siegmund1971} call an ``almost supermartingale'' sequences, and then they apply Robbins \& Siegmund's Theorem~1 to prove that both are vanishing with $P^\star$-probability 1.  But Robbins \& Siegmund also prove a maximal inequality for almost supermartingales and the above proposition's claim is a consequence of their Proposition~2 and Martin \& Tokdar's Theorem~4.8.  We omit the notation-heavy details here.
\end{proof}

This result implies that Martin \& Tokdar's conservative $n^{-(1-\gamma)}$ rate for $K(p^\star, \hat q_{X^n})$ is more accurately the rate associated with the obviously-no-smaller quantity $\max_{s \geq n} K(p^\star, \hat q_{X^s})$.  This is non-trivial because, like supermartingales, almost supermartingale sequences only ``tend'' to decrease.  Unfortunately, it is not clear how this new sense of conservatism in the extant theoretical analyses of $\pr$ can be used to pin down a less conservative bound on rate of convergence.  But the empirical work below may shed some light on this.

\section{Size of the error term in Theorem~2}
\label{s:roc.appendix}

A question posed in the main paper is: how large is the $\pr$e-process's ``$o(n)$'' term in Theorem~2?  The theoretical analysis in \citet{mt-rate} and Section~\ref{A:rate} above imply a conservative bound of $n^\gamma$ on the ``$o(n)$'' term in Theorem~2 of the main paper, when $\pr$'s weight sequence $(w_i)$ satisfies $w_i = (i+1)^{-\gamma}$, for $\gamma \in (2/3,1]$.  So, with the suggested choice of $\gamma \approx 2/3$, a conservative bound on the ``$o(n)$'' term would be roughly $n^{2/3}$.  Less conservatively, away from the edge cases, we expect an improved rate, and the current conjecture is that the $o(n)$ term in question is roughly $n^{\gamma/2}$, or roughly $n^{1/3}$ when $\gamma \approx 2/3$.  In this section, we present some simulation results as empirical evidence in support of this conjecture.  

Towards this, recall the oracle e-process defined in the main paper as
\[ E_n^\text{\sc or} = \frac{p^\star(X^n)}{\sup_{P_0 \in \model_0} p_0(X^n)}, \]
whose magnitude can be used to compare the null model $\model_0$ to the true distribution $P^\star$ based on the data $X^n$.  Then the difference between the log-oracle e-process and the log-$\pr$e-process is 
\[ D_n := \log E_n^\text{\sc or} - \log E_n^\text{\sc pr} = \sum_{i=1}^n \log \frac{p^\star(X_i)}{\hat q_{X^{i-1}}(X_i)}, \]
which is easy to evaluate numerically and does not depend on the null model.  Most important for our present purposes here is that $D_n$ asymptotically agrees with the $o(n)$ term in Theorem~2's bound.  So, if we simulate data and evaluate $D_n$, then we can model $D_n$ as a function of $n$ to approximate its growth and, in turn, estimate the size of the $o(n)$ error term in Theorem~2.  

We simulate 250 data sets for each value of $n$ along the grid $2000, 4000, \ldots, 30000$ and, for each data set, evaluate the difference $D_n$. 
Define $\bar D_n$ to be the average of $D_n$ over the replicates, and then we model $\log \bar D_n$ as a linear function of $\log n$---the coefficient associated with this $\log n$ term represents the polynomial order of $o(n)$, which should be much smaller than 1.  We also fit a model that includes a $\log\log n$ term, but found that the coefficient on this term is not significantly different from zero.  The specific mixture model under consideration here uses a Gaussian kernel $p_u(x) = \nm(x \mid u, 1)$ mixed over the mean in $\UU = [-5,5]$.  For the case when $P^\star$ is in the interior of the convex hull, we take the true mixing distribution to be ${\sf Beta}(3,5)$, centered and scaled to $\UU$.  For the boundary $P^\star$ case, we take the true mixing distribution to be a point mass at 0.  These  experiments were also carried with location--scale normal mixtures and mixtures of gamma kernels and the results were similar to those here. 

\begin{figure}
\centering
\subfigure[Interior-$P^\star$]
{\includegraphics[width = 0.45\linewidth]{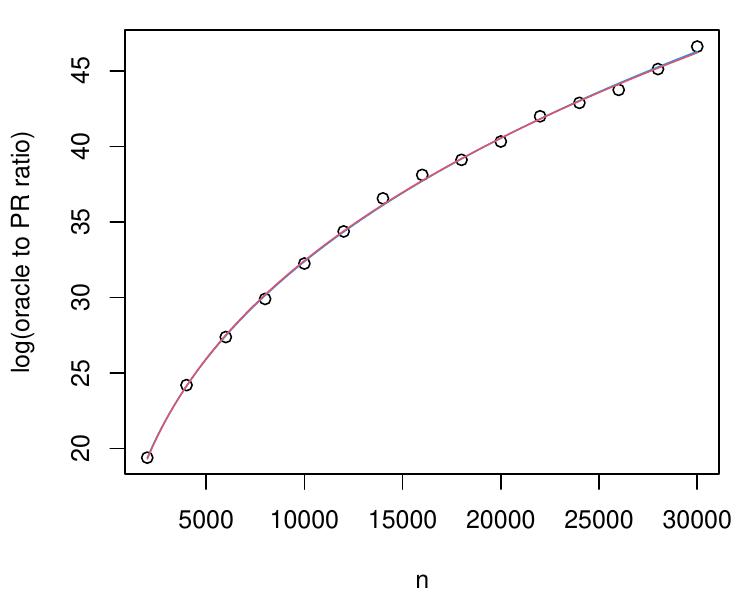}}
\subfigure[Boundary-$P^\star$]
{\includegraphics[width = 0.45\linewidth]{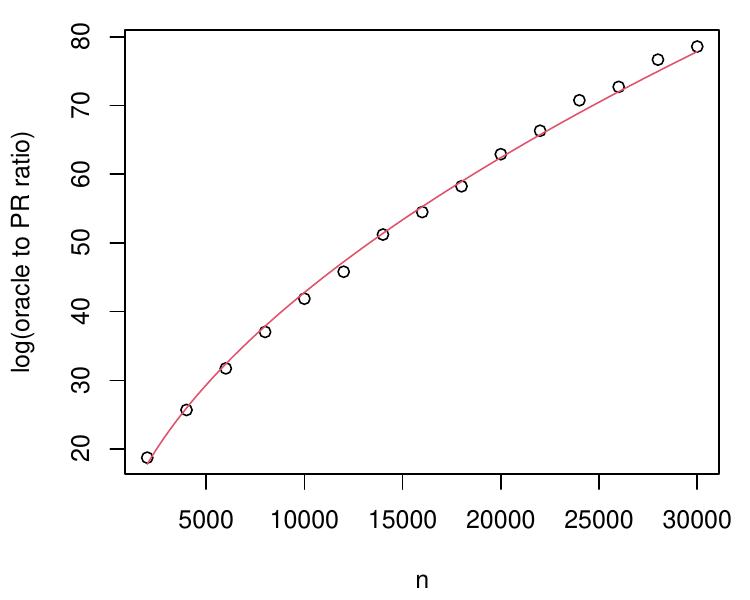}}
\caption{Plots of $\bar D_n$ versus $n$ for two different simulation settings.  In Panel~(a), data are simulated from a $P^\star$ in the interior of the convex hull $\convex$ whereas, in Panel~(b), the data are simulated from $P^\star$ on the boundary of $\convex$.}
\label{fig:rate.sim}
\end{figure}

Figure~\ref{fig:rate.sim} shows our simulation-based estimates of the growth of the $o(n)$ error term as a function of $n$, with Panels~(a) and (b) showing the interior- and boundary-$P^\star$ cases, respectively.  In the interior case, for the linear model $\log \bar D_n \sim \log n$ stated using R syntax, the estimated coefficient on the $\log n$ term is 0.325 with standard error 0.002.  This is in the ballpark of the conjectured value of 0.33 for this interior-$P^\star$ case.  For the boundary-$P^\star$ case, the estimated coefficient is 0.545 with standard error 0.008, which is a little smaller than the conservative bound from \citet{mt-rate}.  

As an aside, the near-2/3 lower bound on $\pr$'s weight sequence parameter $\gamma$ appears to be an artifact of the proof technique employed in \citet{mt-rate}.  One would expect that a smaller value of $\gamma$, e.g., close to the natural lower bound of 0.5, would be a better choice because the slower the weight sequence decays, the more opportunity an individual data point has to influence $\pr$'s estimator.  So, intuitively, one would expect $\pr$ to perform better, to make more efficient use of the data, with smaller choices of the parameter $\gamma$.  If true, then there would be a need for a different strategy for analyzing $\pr$, one that could detect its improved performance with smaller weights.  To check this intuition, and to see if we could get a smaller bound on the $o(n)$ term in Theorem~2, we repeated the above experiment with $\gamma=0.51$.  Contrary to the above intuition, our findings revealed that the aforementioned $o(n)$ term is actually significantly larger for $\gamma=0.51$ than it is for $\gamma=0.67$.  This suggests that there may be something fundamental about the near-2/3 bound in \citet{mt-rate} after all.

\section{Testing log-concavity: multivariate case}

One of the applications presented in the main paper was the use of e-processes, in particular, our proposed $\pr$e-process, for testing log-concavity of the underlying density based on iid observations.  There we focused on a comparison between our $\pr$e-process-based test and the universal inference-based e-process test proposed by \citet{gangrade2023sequential}.  This latter e-process is very similar to the $\pr$e-process: the only difference is that, instead of using $\pr$ to fit a nonparametric mixture density $\hat q_{X^{i-1}}(X_i)$ to each $X_i$ based on $X^{i-1}$, it does so using a standard Gaussian kernel density estimator applied to $X^{i-1}$.  The comparisons in the main paper focused on data of dimension $d=1$.  Our goal here is to do a similar comparison for $d > 1$.  

First, we need to comment on the computation of the two e-processes when the dimension is greater than 1.  Lets start with the denominators.  As in the main paper, for dimension $d=1$, the log-concave maximum likelihood can be found using the {\tt logConDens} function from the R package 
{\tt logcondens} \citep{dumbgen2011logcondens}; for $d \in \{2,3,4\}$ this can be done using the R package at \url{https://github.com/FabianRathke/fmlogcondens}.  Next, we discuss computation of the e-process numerators in turn.  
\begin{itemize}
\item For the kernel-based universal inference, we obtain the $d$-dimensional kernel density estimate using the {\tt kde} function in the R package {\tt ks} \citep{duong_ks.R}; the same choice is made in \citet{gangrade2023sequential}.  This can accommodate dimension $d \leq 6$.  We use a diagonal, data-driven bandwidth matrix as implemented in the {\tt Hpi.diag} function in the same R package.  Doing this sequentially, i.e., constructing a kernel density estimator at $X_i$ based on $X^{i-1}$ for each $i \geq 1$, is quite expensive and, for this reason, \citet{gangrade2023sequential} propose a batching scheme.  Here we are more interested in the quality of the e-processes than in the computational efficiency, so we did not implement their batched version. 
\item For $\pr$, implementation is easy when the dimension is low.  Here, however, for $d$-dimensional data, to fit a multivariate normal mixture, where mixing is over the mean vector and diagonal covariance matrix, then the dimension of the latent $\UU$-space is $2d$.  This means we quickly surpass cases where $\pr$'s integration can be done using simple quadrature rules.  It is for precisely these cases that we developed the $\pr$ticle filter strategy in \citet{prticle}, which goes roughly as follows.  Start with a random sample $\{U_t: t=1,\ldots,T\}$ from $\pr$'s initial guess $\hat\Psi_0$ of the mixing distribution supported on $\UU$ and, to each $U_t$, attach the weight $\pi_0(U_t) \equiv 1$.  As described in \citet[][Sec.~3]{prticle}, the weights are sequentially updated as data points arrive so that the weighted samples after observing $X^{i-1}$ can be used to approximate expectations with respect to the $\pr$ estimator $\hat\Psi_{i-1}$.  In particular, they propose 
\[ \tilde q_{X^{i-1}}(X_i) := \frac1T \sum_{t=1}^T k(X_i \mid U_t) \, \pi_{i-1}(U_t) \]
as an approximation of $\pr$'s $\hat q_{X^{i-1}}(X_i) = \int k(X_i \mid u) \, \Psi_{i-1}(du)$.  Importantly, the weight updating scheme requires no integration.  With this approximation of $\pr$'s predictive density, it is straightforward to get the corresponding $\pr$e-process.  
\end{itemize} 

To compare the two e-processes, we follow the strategy in Section~\ref{s:roc.appendix} above.  That is, we compare each of the e-processes to the oracle e-process that knows the true density $p^\star$.  In particular, we consider 
\[ D_n^\text{\sc pr} = \log E_n^\text{\sc or} - \log E_n^\text{\sc pr} \quad \text{and} \quad D_n^\text{\sc ui} = \log E_n^\text{\sc or} - \log E_n^\text{\sc ui}, \]
where $E_n^\text{\sc pr}$ and $E_n^\text{\sc ui}$ are the $\pr$e-process and kernel-based universal inference e-process, respectively.  The advantage of this comparison is that it alleviates the computation of the log-concave maximum likelihood estimators and focuses specifically on the aspect that distinguishes the two e-processes, namely, their choice of numerator.  

We focus our comparison on the case of $d=3$.  Since the kernel density estimation scheme can only handle $d \leq 6$, we consider the choice $d=3$ to be ``moderate dimension.''  In this case, implementation of $\pr$---fitting a Gaussian mixture model with diagonal covariance matrix---would require integration over a 6-dimensional space, which cannot be done using quadrature.  Here, instead, we use the $\pr$ticle filter approximation described above with $T=20,000$ initial particles.  This can all be done up to at least dimension $d=5$, but all the computations (both kernel and $\pr$) become more expensive.  Since the point we want to make is apparent with $d=3$, this is where we focus our investigation.  

\begin{figure}
\centering
\includegraphics[width = 0.6\linewidth]{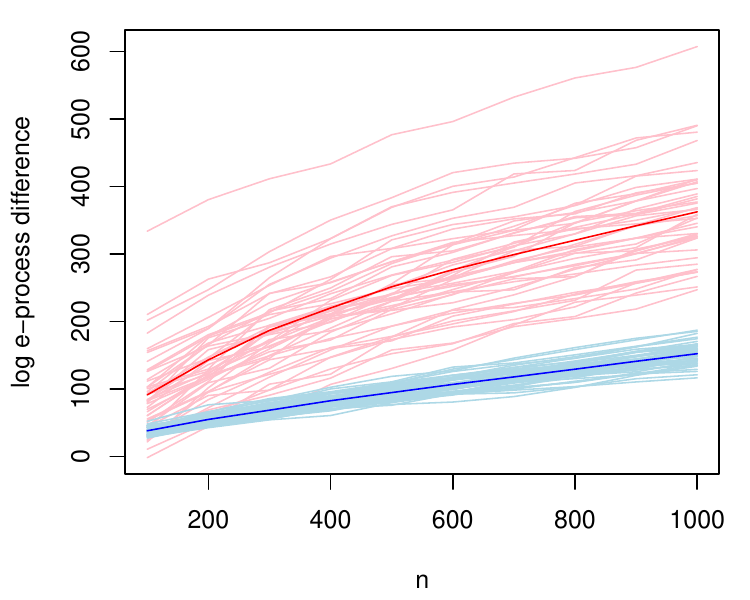}
\caption{Plots of $D_n^\text{\sc pr}$ (blue) and $D_n^\text{\sc ui}$ (pink) versus $n$ for 50 data sets, along with the pointwise average.}
\label{fig:dim.sim}
\end{figure}

Figure~\ref{fig:dim.sim} shows the paths $n \mapsto D_n^\square$ for the two e-processes, $\square \in \{\pr, \text{\sc ui}\}$, under investigation over 50 data sets sampled iid from a true distribution $P^\star$ that is a two-component mixture of 3-dimensional Gaussian kernels with different means and different diagonal covariance matrices.  This means that, in a certain sense, both e-processes are fitting a ``correctly-specified model'' in their respective numerators, just with a different estimator: $\pr$ versus a kernel density estimator.  Since $D_n^\square$ measures the gap between the given e-process and the idealized oracle e-process, smaller values of $D_n^\square$ indicate a smaller gap and, hence, a better approximation of that ideal.  As is clear from the plot, $D_n^\text{\sc pr}$ tends to be significantly smaller than $D_n^\text{\sc ui}$ and far less variable across replications.  The take-away message is that, while the increased dimension is sure to affect the quality of an e-process's approximation of the oracle e-process, but the increased dimension has a more significant effect on the kernel-based e-process than it does on the $\pr$e-process implemented with the $\pr$ticle filter.  Moreover, although we did not implement the batch procedure suggested in \citet{gangrade2023sequential}, it is worth mentioning that the $\pr$e-process's denominator is much less expensive to compute than the kernel-based e-process's, despite the fact that the latter is working with a relatively large number of particles.  This is because the $\pr$ updates are genuinely recursive while a naive implementation of the kernel update, with data-driven bandwidth, requires a look-back at the full data.  

\section{An alternative PR-like e-process}

While there are a number of upsides to the use of a mixture model and $\pr$ in the denominator of the e-process construction, there are also downsides.  One is that it is required to specify the mixture model's kernel $\{p_u: u \in \UU\}$.  In some cases, this specification is trivial, i.e., one might just adopt a mixture of normals for the sake of flexibility, but in other cases it might not be.  A natural question is, if all that one needs is the sequence of one-step-ahead predictive densities at the observations, then why is the mixture model formulation necessary?  Can we not just work directly with densities on the sample space?  \citet{hahn.martin.walker.pred} considered this question in the context of recursive Bayesian updating, and a proposal they put forward is the following.  Their primary focus was on scalar observations so that is the context we consider here.  Like with $\pr$, start with a sequence of weights $(w_i)$ and a prior (univariate) predictive density $q_0$ with corresponding distribution function $Q_0$.  For a sequence of observations $X_1,X_2,\ldots$, they suggest the following sequence of predictive densities 
\[ q_n(x) = (1-w_n) \, q_{n-1}(x) + w_n \, q_{n-1}(x) \, c_\rho\bigl( Q_{n-1}(x), Q_{n-1}(X_n) \bigr), \quad x \in \XX, \quad n \geq 1, \]
where $c_\rho$ is the Gaussian copula density, 
\[ c_\rho(u, v) = \frac{\nm_2( \Phi^{-1}(u), \Phi^{-1}(v) \mid 0, 1, \rho)}{\nm(\Phi^{-1}(u) \mid 0, 1) \, \nm(\Phi^{-1}(v) \mid 0, 1)}, \]
with $\Phi$ the standard normal distribution function, and $\nm_2$ in the numerator the bivariate normal density, with zero means and unit variances, depending on a correlation parameter $\rho \in [0,1)$.  Again, the advantage here compared to the original $\pr$ is that it does not require specification of a mixture model or the numerical integration to convert $\pr$'s mixing distribution estimator into a corresponding mixture density estimator.  The recursion is written most naturally in terms of distribution---rather than density---functions, but then numerical differentiation is needed to get the density for the likelihood calculations that follow.  

From the predictive updates, one can get a corresponding joint density via multiplication:
\[ q^\text{\sc rbp}(X^n) = \prod_{i=1}^n q_{i-1}(X_i), \]
where {\sc rbp} indicates that this is motivated by recursive Bayes predictive updating.  Given a null hypothesis $\model_0$, one can define the corresponding $\pr$-like e-process as 
\[ E_n^\text{\sc rbp} = \frac{q^\text{\sc rbp}(X^n)}{\sup_{p_0 \in \model_0} p_0(X^n)}. \]

This alternative e-process is interesting, and here are some points that we think warrant further investigation.  First, there is currently no theory available that can be used to investigate the asymptotic growth rate of $E_n^\text{\sc rbp}$, so can anything be said about this?  Second, how does this perform empirically compared to the $\pr$e-process and other available e-processes?  Third, can this be extended naturally---and in a way that can be theoretically analyzed---to handle multivariate observations?  A suggestion for handling bivariate data is given in \citet{hahn.martin.walker.pred}, along with some brief numerical results, but no general formulation or theory.

\end{document}